# Chirality sensing employing Parity-Time-symmetric and other resonant gain-loss optical systems

Ioannis Katsantonis[1,2*], Sotiris Droulias[1,5], Costas M. Soukoulis[1,3], Eleftherios N. Economou[1,4], T. Peter Rakitzis[1,4] and Maria Kafesaki[1,2‡]

[1]*Institute of Electronic Structure and Laser, Foundation of Research and Technology Hellas, 71110, Heraklion, Crete, Greece*
[2]*University of Crete, Department of Material Science and Technology, 71003, Heraklion, Greece*
[3]*Ames Laboratory and Department of Physics and Astronomy, Iowa States, Ames Iowa, 50010, USA*
[4]*University of Crete, Department of Physics, 71003, Heraklion, Greece*
[5]*Department of Digital Systems, University of Piraeus, Piraeus, 18534, Greece*

Molecular chirality detection and enantiomer discrimination are very important issues for many areas of science and technology, prompting intensive investigations via optical methods. However, these methods are hindered by the intrinsically weak nature of chiro-optical signals. Here, we investigate and demonstrate the potential of gain materials and of combined gain-loss media to enhance these signals. Specifically, we show that the proper combination of a thin chiral layer with a gain-loss bilayer can lead to large enhancements of both the circular dichroism (CD) response and the dissymmetry factor, $g$, compared to the chiral layer alone. The most pronounced enhancements are obtained in the case of a Parity-Time (*PT*) symmetric gain-loss bilayer, while deviations from the exact *PT* symmetry lead to only moderate changes of the CD and $g$ response, demonstrating also the possibility of tuning the system response by tuning the gain layer properties. In the case of *PT*-symmetric gain-loss bilayers we found that the largest CD enhancement is obtained at the system lasing threshold, while the $g$-enhancements at the anisotropic transmission resonances of the systems. Our results clearly demonstrate the potential of gain materials in chirality detection. Moreover, our gain-involving approach can be applied in conjunction with most of the nanophotonics/nanostructures-based approaches that have been already proposed for chirality sensing, further enhancing the performance/output of both approaches.

## I. INTRODUCTION

Chiral objects, i.e. three-dimensional objects that present mirror asymmetry [1], are all around us, ranging from our DNA and other important biomolecules to chemical drugs, and extending over spiral galaxies [2]. *Chiral objects* are classified according to their handedness as left-handed or right-handed, with the left- and right-handed forms of an object known as enantiomers. There are many chiral biomolecules and chemical drugs for which the two enantiomers interact differently with a biological organism having severely different therapeutic and/or toxicological effects [3]. Therefore, an efficient detection and discrimination of the different enantiomers of a chiral substance is crucially important in many scientific fields, like medicine, pharmacology, biology, chemistry [4] and fundamental physics [5]. A major problem though in such a detection is the very weak chiro-optical signals encountered in all the light-related schemes applied for chirality detection.

Chirality, besides being a property of matter is also a property of the electromagnetic field [6]. Although the electromagnetic field cannot be chiral in the conventional sense, as it is not an object, the chirality arises from the polarization rotating electric and magnetic fields as the wave propagates in space, forming left- or right-handed helical patterns. The field chirality is usually quantified by the quantity $C = -\omega \text{Im}[\boldsymbol{E} \cdot \boldsymbol{H}^*]/2c^2$, proportional to the inner product of electric and magnetic field, which is called optical chirality [6,21] ($\omega$ is the wave frequency and $c$ the vacuum light speed). The simplest example of a chiral field is circularly polarized (CP) light, and it has been used for over 200 years to measure molecular chirality [7]. Fields of higher $C$ than CP light are characterized as superchiral. Recent studies have shown that superchiral fields can play an important role in the detection of chiral molecules, as they offer enhancement of chiro-optical signals, greatly facilitating thus the chirality detection and the discrimination of the different enantiomers [6]. Therefore, many studies have been devoted recently in the generation of strong superchiral fields and their exploitation for chiral molecule detection [8-21]. Many of them have explored various nanophotonic [9], plasmonic [10,11] and metamaterial platforms [14] as means to enhance chiral local fields. For example, high-index dielectric nanoparticles [9,13,15-17], plasmonic spheres [8] or disks [9], helicity preserving optical cavities [12,18,19,22-24] and others [15,25-27] have been employed for chiral detection and enhancement of chiral sensing. However, all the above mentioned approaches target enhancement of the circular dichroism (CD) response signal (i.e. the absorption difference between right-handed circularly polarized (RCP/+) and left-handed circularly polarized (LCP/-) incident wave, i.e. $A_+ - A_-$), a quantity directly proportional to the volume of the chiral substance to be detected. A quantity strongly related with CD, of additional merit though regarding chiral interactions and chirality sensing, is the so-called Kuhn's dissymmetry factor, $g$ [17,21,28-30], defined as the absorption difference for RCP and LCP waves (i.e. the CD) divided by the average absorption (i.e. $(A_+ + A_-)/2$):

$$g = \frac{(A_+ - A_-)}{(A_+ + A_-)/2} \propto \frac{8c\kappa''C}{\omega(\varepsilon''|\boldsymbol{E}|^2 + \mu''|\boldsymbol{H}|^2)} \quad (1)$$

In (1) $C$ is the optical chirality (see appendix F for details) and $\varepsilon''$, $\mu''$, $\kappa''$ the imaginary part of electric permittivity, magnetic permeability and chirality parameter, respectively, of the chiral medium. (For the derivation of the r.h.s of (1) see Appendix F.)

Dissymmetry factor, $g$, is a useful dimensionless quantity describing the relative preferential absorption of circularly polarized waves by a chiral sample. Enhanced $g$ results to enhanced enantio-selectivity in all chiral light-matter interactions involving light absorption, including fluorescence (used for, e.g. sensing [21]), photolysis, photo-polymerization [4] etc.). From Eq. (1) it is evident that the relative chiral asymmetry in the absorption is proportional to both the chirality of matter ($\kappa$) and the chirality of the electromagnetic field ($C$); therefore its maximization through proper electromagnetic field enhancement or structuring promises a viable route towards enhanced enantio-selectivity in chiral light-matter interactions. Maximization of $g$ though using common nanophotonic resonances and platforms [21,26,31,32] is not a straightforward task, as the maximization of the absorption difference there (numerator of $g$) is always associated with enhancement of the absolute absorptions $A_+$ and $A_-$ ($g$ denominator), which reduces the $g$ value. Approaches to enhance $g$, besides the change/optimization of the nanophotonic or metamaterial structure employed, include optical approaches to create super-chiral light. The dominant such approach is the exploitation of field nodes created in a cavity by the interference of two counter-propagating CP waves [21], while other approaches include an efficient manipulation of the orbital angular momentum delivered by optical vortices [33], use of non-linear phenomena [34], etc. However, most of the above mentioned approaches, especially the first one, are associated with small field intensities in the region of the superchiral fields, limiting thus the overall efficiency of the light-matter interaction.

A simple approach that can decouple the simultaneous maximization of numerator and denominator of $g$, being thus suitable for its maximization without concurrent minimization of the overall field intensity, can be offered by the involvement of gain media. Here we propose and investigate such an approach and its potential and capabilities for circular dichroism and dissymmetry factor enhancement in a molecular chirality sensing scheme. In particular, we propose a simple structure combining a thin chiral layer (which is the chiral material under detection) with gain and loss media in balance, as to exhibit parity-time (*PT*-) symmetry, or beyond such balance. *PT*-symmetric systems exploit strong electromagnetic fields at the interfaces of the gain-loss media [35,36] as well as exotic scattering features, such as anisotropic transmission resonances (ATRs) [35], and exceptional points associated with strong sensitivity [37,38,52,53]. Due to such peculiar features *PT*-symmetric optical systems have gained a growing attention in both fundamental and applied research [39-53].

As already mentioned, there are many works investigating and proposing schemes for efficient molecular chirality detection and enantiomer discrimination. Since chiral light-matter interactions are inherently extremely weak (at optical frequencies, natural materials have $\kappa \sim 10^{-5}$), this detection can be very challenging, especially when only tiny amounts of substances are involved, e.g. ultrathin chiral layers. We show here that our approach of combining a thin chiral layer with a gain-loss *PT*-symmetric bilayer can to a large degree overcome the main challenges involved in the chiral detection. More specifically, by both analytical and numerical calculations, we demonstrate strong circular dichroism (CD) signals in our *PT*-symmetric system and high values of the dissymmetry factor $g$. Furthermore, we investigate the necessity of the *PT*-balanced gain-loss in our system; by changing the gain-loss ratio we are still able to achieve enhanced circular dichroism and dissymmetry factor, while overcoming the strict *PT*-symmetry highly facilitates experimental realization and validation of our findings. Investigating further the origin and the potential of the $g$ enhancement in our systems, we find that: (a) in *PT*-symmetric systems it occurs at the anisotropic transmission resonance (ATR) points of the systems [35], which are associated with unidirectional zero reflection and unity transmission; (b) $g$-enhancement is associated with no-reduction or even enhancement of the total field intensity, a quantity critical for the overall efficiency of any light-matter interaction. Finally, we have to mention that our approach of employing gain media for chirality sensing can be used in conjunction with most of the up to now proposed nanophotonics-based approaches, expanding further their potential as chirality sensing platforms, and proposing the area of active nanophotonics as a viable route for chirality sensing and differentiation.

## II. PT-SYMMETRIC TRILAYERS

In our study we first consider a three layer geometry with total thickness $L = d + d_c + d$ along *z*-direction (and infinite along *x*- and *y*-directions) in free space, as shown in Fig. 1, where a thin chiral layer, of thickness $d_c$, is sandwiched between a gain and a loss layer, both of thickness *d*. We use the $\exp(-i\omega t)$ time convention (in this convention the positive imaginary part of the refractive index describes loss, while negative values of this quantity correspond to gain).

We are interested in exploiting the electromagnetic response of our purposed system under circularly polarized incident waves. We solve Maxwell's equations $\nabla \times \boldsymbol{E} = i\omega \boldsymbol{B}$ and $\nabla \times \boldsymbol{H} = -i\omega \boldsymbol{D}$ with the appropriate boundary conditions and for the chiral layer we assume the constitutive relations [54] $\boldsymbol{D} = \varepsilon_r \varepsilon_0 \boldsymbol{E} + i(\kappa/c)\boldsymbol{H}$ and $\boldsymbol{B} = \mu_r \mu_0 \boldsymbol{H} - i(\kappa/c)\boldsymbol{E}$, where $\varepsilon_r, \mu_r, \kappa$ refer to the relative permittivity, permeability and the chirality (Pasteur) parameter (quantifying the magnetoelectric coupling), respectively ($\varepsilon_0, \mu_0$ are the vacuum permittivity and permeability, respectively). Inserting the above constitutive relations into Maxwell's equations, we can find the chiral Helmholtz equation to be $\nabla^2 \boldsymbol{E} + \omega^2 \left\{\frac{(\varepsilon_r \mu_r - \kappa^2)}{c^2}\right\} \boldsymbol{E} + \left[\frac{2\omega\kappa}{c^2}\right] \nabla \times \boldsymbol{E} = 0$. Assuming plane waves propagating along *z*, we obtain the elementary solutions of the chiral Helmholtz equation as $\boldsymbol{E}_\pm = E_\pm(\hat{\boldsymbol{x}} \pm i\hat{\boldsymbol{y}})e^{i(k_\pm z - \omega t)}$, where $k_{+/-} = \omega(\sqrt{\varepsilon_r \mu_r} \pm \kappa)/c$ are the wave vectors in the chiral slab and $E_+, E_-$ are the amplitudes for right- and left-handed circularly polarized waves, respectively (employing source side view). The magnetic field can be found from $\boldsymbol{E}_\pm$ by $\boldsymbol{H}_\pm = \mp i Z^{-1} \boldsymbol{E}_\pm$, where $Z = \sqrt{\frac{\mu_r \mu_0}{\varepsilon_r \varepsilon_0}}$ is the wave impedance.

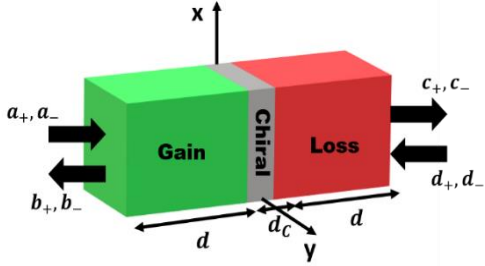

Figure 1. A three layer system, consisting of a thin layer of a chiral medium sandwiched between a gain and a loss slab (both of thickness $d$ along $z$-direction and infinite along $x$ and $y$). $a_\pm$, $b_\pm$, $c_\pm$ and $d_\pm$ are the amplitudes of ingoing and outgoing (see arrows) RCP/+ and LCP/- waves.

For the (non-chiral) gain and loss media the above equations are simplified by taking $\kappa = 0$, so the wave vector takes the well-known form $k = \frac{\omega}{c}\sqrt{\varepsilon_r \mu_r}$. To obtain general expressions allowing investigation of different material combinations we start with layers of arbitrary material parameters. Considering incident left- and right-handed circularly polarized waves and requiring the continuity of the tangential components $(x,y)$ of $\boldsymbol{E}$ and $\boldsymbol{H}$ at each of the four interfaces of the system $\left(z = -\left(d + \frac{d_C}{2}\right), z = -\frac{d_C}{2}, z = \frac{d_C}{2} \text{ and } z = \left(d + \frac{d_C}{2}\right)\right)$, we obtain a 16x16 linear system of equations which is solved analytically, giving the scattering coefficients (transmission and reflection) of the total system (see Appendix A for analytical expressions).

### III. CIRCULAR DICHROISM

In order to characterize the chiral response of our system, and through it the chiral properties of the chiral layer under investigation, we further analyse the scattering properties, calculating the circular dichroism (CD), $CD = A_+ - A_-$. The absorption coefficients are given by $A_+ = 1 - |t_{++}|^2 - |r_{-+}|^2$ for RCP/+ and $A_- = 1 - |t_{--}|^2 - |r_{+-}|^2$ for LCP/- incidence, where the first subscript in the transmission ($t$) and reflection ($r$) coefficients indicates the output polarization and the second the incident polarization. The different subscripts in the reflection coefficients are due to the property of chiral interfaces to reverse the circular polarization upon reflection [55,56]; i.e., when an incidence wave is RCP, it is transformed to LCP upon reflection and vice versa. By analytical calculations, we find $r_{-+} = r_{+-}$ for incidence from either side of the system; hence the circular dichroism is proportional to the transmission difference for LCP and RCP waves and it is the same for left-side and right-side incident waves due to reciprocity. Calculating the CD for the particular case of $PT$-symmetric loss and gain slabs, i.e. with equal real part of permittivity and permeability and opposite imaginary parts [35-40], we find

$$CD = |t_{--}|^2 - |t_{++}|^2 = \frac{512 e^{-2kd_C \text{Im}[n_C]}}{|Z_G Z_C Z_L (A_1 + A_2 + B_1 + B_2)|^2} \sinh(2kd_C \text{Im}[\kappa]), \quad (2)$$

where $k = \omega/c$ is the wave number in the vacuum and $n_C = \sqrt{\varepsilon_C \mu_C}$ is the non-chiral part of the index of refraction in the chiral medium. The wave impedance in each layer is $Z_i = \sqrt{\mu_0 \mu_i / \varepsilon_0 \varepsilon_i}$, where the subscript $i = \{G, C, L\}$ denotes the *gain*, *chiral* or *loss* region, respectively. The terms $A_1, A_2$ and $B_1, B_2$ in Eq. (2) depend on the impedances and wave vectors and are totally independent of the chirality parameter $\kappa$ (see appendix A and B). Thus, Eq. (2) shows that circular dichroism depends only on the imaginary part of the chirality parameter (through $\sinh(2kd_C \text{Im}[\kappa])$; for positive chirality, $\text{Im}[\kappa] > 0$, the hyperbolic sin is positive while for negative chirality, $\text{Im}[\kappa] < 0$, sinh is negative; this change of sign allows employment of the system for discrimination of the two different enantiomers of a chiral molecular system (note that the two enantiomers have opposite chirality parameter, $\kappa$). The sinh prefactor in Eq. (2) depends only on the non-chiral absorption of the chiral layer (via $\text{Im}[n_c]$) as well as on the gain-loss media, and determines the position of the resonances of the total system.

In the following we apply our formalism to systems with close to realistic material parameters, in order to investigate and quantify the potential of those systems for molecular chirality sensing. In Fig. 2 (a), we plot the circular dichroism (CD) for a thin chiral layer alone (of thickness $d_C$=10 nm), with chirality parameter $\kappa = \pm 5(10^{-4} + 10^{-5}i)$, which is very close to the chirality of aqueous solutions of chiral molecules [22] and with a typical ratio between real and imaginary parts [57]; its non-chiral refractive index is $n_C = 1.33 + 0.01i$. In Fig. 2(b), we plot the CD for the thin chiral slab placed in the middle of a $PT$-symmetric bi-layer, as shown in Fig. 1 (see also Fig. 8, Appendix D for the CD in a wider frequency range). The gain and loss slabs have refractive index $n_{G/L} = 3 \mp 0.04i$ where the (-) sign corresponds to the gain medium and the (+) sign to the loss medium, as dictated by $PT$-symmetry [46]. (Note that $PT$-symmetric optical systems with parameters very close to ours have already been studied both theoretically [46] and experimentally [58,59]. Note also that the gain value employed, although not achievable with the current bulk gain materials, can be achieved (as effective gain) in gain-involving metamaterials, due to the resonance-induced enhancement. Each of the two layers has thickness $d = 2.5\ \mu m$. The black lines in Fig. 2 correspond to positive sign in the chirality parameter and the red lines to negative sign. We observe that the CD depends on the sign of the imaginary part of chirality, as predicted from Eq. (2). Comparing Figs. 2(a) and 2(b), we observe that the presence of the $PT$-symmetric bilayer leads to circular dichroism up to seventy five (75x) times larger than that of the chiral layer alone. As a comparison, we note that the CD signal of our "$PT$-symmetric" configuration is close to one order of magnitude larger than the CD signals of the chiral and achiral nanophotonic systems that have been already studied and proposed for chirality sensing [8].

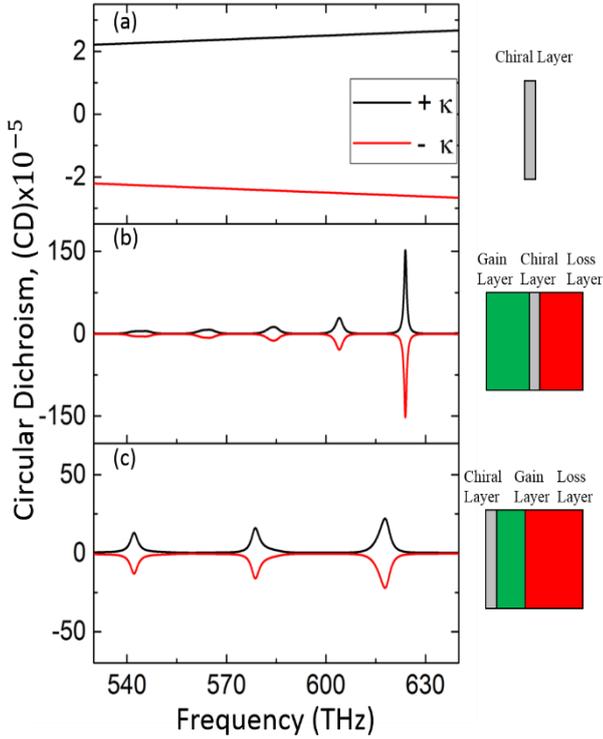

Figure 2. Circular dichroism (CD) for the system shown in Fig. 1 and two related systems. Panel (a) corresponds to a chiral layer alone; panel (b) corresponds to the *PT*-symmetric system shown in Fig. 1 with the middle/chiral layer the same as in panel (a); panel (c) corresponds to a general non-Hermitian system, again with the same chiral layer (gray area) but placed next to the gain layer. For geometrical and material parameters see main text. The black lines correspond to positive sign of the chirality (both real and imaginary), $\kappa$, of the chiral layer, while the red lines to negative sign. The transmission amplitudes employed for the CD calculation, as well as the CD in a broader frequency region are shown in Fig. 8, Appendix D.

Because of the strict conditions imposed by the full "*PT*-symmetry" requirement, which limit the practical realization and exploitation of such systems, it is important to know if and to what extent the exact *PT*-symmetry condition is essential for the observed CD enhancement. For that we investigated also general non-Hermitian systems where the gain-loss media are not in balance. An example is shown in Fig. 2 (c) (see also Fig. 8, Appendix D) where we assume a gain layer with index $n_G = 2 - 0.05i$ and thickness $d_G = 2~\mu m$, a loss layer with index $n_L = 3 + 0.04i$ and thickness $d_L = 3~\mu m$ and where the chiral layer (same as in Figs. 2(a), (b)) is placed at the left side of the gain-loss bilayer (i.e. it is attached to the gain layer); this configuration is more amenable to experimental realization. We observe also here a comparable to Fig. 2(b) CD enhancement, despite the thinner-gain and thicker-loss layer. In particular, we observe that at frequencies close to 617 THz, we have values of CD more than ten times larger than that of the chiral layer alone. The large CD enhancement regions in both the cases of Figs. 2(b) and 2(c) coincide with the transmission resonances of the structures (Fabry-Perot resonances; see Fig. 8, Appendix D), while the largest enhancement is obtained at frequencies 643.7 THz and 655 THz, respectively, which are close to the lasing threshold of the corresponding structure [35] – see Fig. 8 in Appendix D,

Note that the CD peaks of both systems of Fig. 2 are associated with quite high quality-factor ($Q = \frac{\text{Re}[\omega_0]}{2\text{Im}[\omega_0]} >$ 100, where $\omega_0$ is the resonance frequency) not only around the lasing threshold, but also below lasing threshold (and exceptional point). This is attributed to the balanced gain-loss, in particular in the *PT*-symmetric case. In *PT*-symmetric systems below exceptional point the gain balances the loss [35-48] and we have almost lossless modes. As the frequency increases and the system approaches the lasing threshold a transmission pole in the complex frequency plane approaches and finally crosses the real axis, i.e. $\text{Im}[\omega_0] \to 0 \Rightarrow Q \to \infty$.

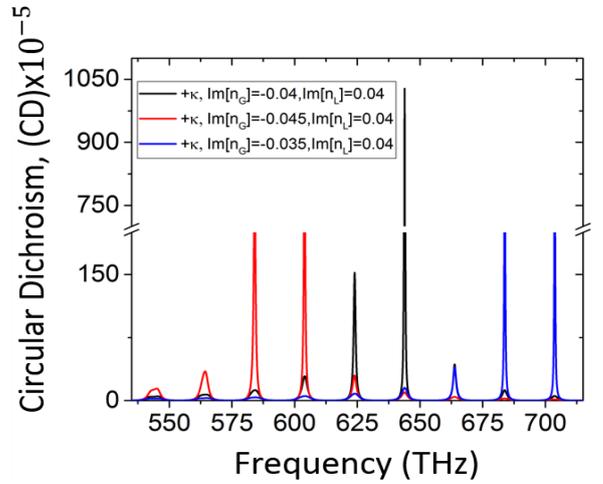

Figure 3. Circular dichroism (CD) three different gain-loss [Im($n_G$)-Im($n_L$)] ratios for the system of Fig. 2 (b). In all cases we assume positive sign of the chirality parameter (both real and imaginary parts) of the thin chiral layer.

To investigate further and understand the role and necessity of *PT*-symmetry regarding the achievable CD enhancement, we calculated the CD for the system of Fig. 1 for different gain-loss contrasts. The result is shown in Fig. 3, and compared with the exact *PT*-case (the one of Fig. 2(b) and Fig. 8 of Appendix D). As can be seen in Fig. 3, a small deviation (increase/decrease of gain) from the exact *PT*-symmetric case (black line in Fig. 3) leads to only moderate deterioration of the CD peaks. Moreover, with increase of the gain parameter (Im($n_G$)), we observe a frequency shift in the maximum CD response due to lasing threshold and the corresponding "exceptional point" moving in lower frequencies [35], while the opposite behavior is observed with decrease of the gain parameter. We have to note here than when the system reaches lasing threshold it becomes unstable (a self-sustained oscillator), hence, quantities such as transmittance power are not well defined. That's why we concentrate our attention mostly before lasing threshold, where also strong CD enhancement takes place. The results of Fig. 3 clearly demonstrate that *PT*-symmetric gain-loss is a sufficient but not a necessary condition for large CD enhancement in gain-loss systems.

## IV. DISSYMMETRY FACTOR

As already mentioned in the introduction, dissymmetry factor determines to a large degree the percentage of enantio-selectivity in all photo-induced chiral light-matter interactions (photoionization, fluorescence, photolysis, etc.), which can be exploited not only for enantiomer discrimination but also for enantioselective synthesis, critical for the pharmaceutical industry. To evaluate the dissymmetry factor in our system we employ the defining Eq. (1); there we notice that the numerator, CD (see Eq. (2)), depends on the chirality as $\sinh(2kd_C \text{Im}[\kappa])$ and it is independent of the side of incidence. The total absorption though (denominator of Eq. (1)), $A_+ + A_- = 2 - |t_{++}|^2 - |t_{--}|^2 - |r_{-+}|^2 - |r_{+-}|^2$, depends on the reflection which is different for waves incident from opposite system sides, making the dissymmetry factor side-dependent. From analytical calculations we find for the system of $PT$-symmetric gain-loss layers the dissymmetry factor for waves incident from the left, $g^{(L)}$, and from the right, $g^{(R)}$, as

$$g^{(L)} = -2\tanh(2kd_C\text{Im}[\kappa]) + 512e^{-2kd_C\text{Im}[n_C]}\sinh(2kd_C\text{Im}[\kappa])\left[\frac{1}{P} + \frac{1}{Q^{(L)}}\right], \quad (3)$$

$$g^{(R)} = -2\tanh(2kd_C\text{Im}[\kappa]) + 512e^{-2kd_C\text{Im}[n_C]}\sinh(2kd_C\text{Im}[\kappa])\left[\frac{1}{P} + \frac{1}{Q^{(R)}}\right], \quad (4)$$

where $P = |Z_G Z_C Z_L (A_1 + A_2 + B_1 + B_2)|^2$, $Q^{(L)} = |Z_G Z_C Z_L (C_1^{(L)} + C_2^{(L)} + D_1^{(L)} + D_2^{(L)})|^2$ and $Q^{(R)} = |Z_G Z_C Z_L (C_1^{(R)} + C_2^{(R)} + D_1^{(R)} + D_2^{(R)})|^2$. The terms $C_1^{(R)}, C_2^{(R)}, D_1^{(R)}, D_2^{(R)}$ in $Q^{(R)}$ and $C_1^{(L)}, C_2^{(L)}, D_1^{(L)}, D_2^{(L)}$ in $Q^{(L)}$ are given in appendix A.

To obtain quantitative data regarding the dissymmetry factor, we consider the same systems as in Fig. 2. In Fig. 4 we plot the dissymmetry factor, $g$, for circularly polarized waves impinging our systems from both sides. In panels (a) and (d), we show the $g$ values for the chiral layer alone (for comparison); it is evident that $g$ depends on the sign of chirality (through the CD - numerator of Eq. (1)). In panels (b) and (e), we present the $g$ values for the "$PT$-symmetric" case. We observe two discrete side-dependent peaks which demonstrate an up to 7x $g$-enhancement compared to the chiral layer alone. The positions of the peaks are at frequencies where the total absorption (denominator of Eq. (1)) approaches zero (but not exactly zero, where $g$ goes to infinity), as we will discuss and analyze in more detail later on. Finally, in panels (c) and (f), we observe similar to the "$PT$-symmetric" case behavior in the more general non–Hermitian system of "unbalanced" loss and gain materials (system of Fig. 2(c)). Again here, the $g$-peaks are observed at frequencies where the total absorption (denominator of $g$) tends to zero. (Note that although the term "absorption" is usually connected with material losses and takes positive values, here we keep the same terminology even for materials with gain, meaning field enhancement and taking negative values.) Observing the results of Fig. 4, one can see also an asymmetry in the g-resonances profile, resembling anti-symmetric Fano profiles, similar to the ones discussed in Ref. [60]. This asymmetry here seems to originate from the asymmetric coupling of the chiral layer with the gain and the loss layers, as well as from the asymmetry of the propagation (and absorption) characteristics for left and right incidence, as is revealed from the corresponding field analysis (see Section VI).

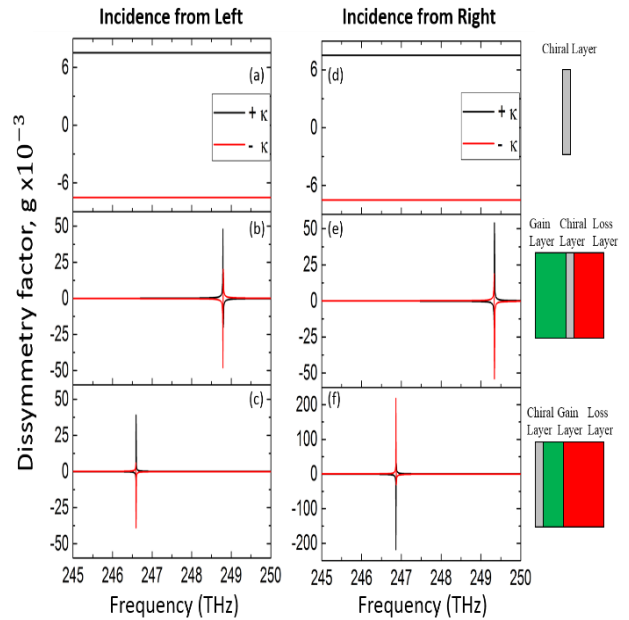

Figure 4. Dissymmetry factor, $g$, calculations for the three systems of Fig 2, with waves incident from the left side (left column) and waves incident from the right side (right column) of the systems; panels (a) and (d) correspond to the chiral layer alone, panels (b) and (e) correspond to the "$PT$-symmetric" system and panels (c) and (f) to the general non-Hermitian system. The black lines correspond to positive sign of the chirality (both real and imaginary), $\kappa$, of the chiral layer, while the red lines to negative sign. The material parameters are the same as in Fig. 2.

## V. SCATTERING MATRIX ANALYSIS OF THE PT-TRILAYER STRUCTURE

To fully understand the response of our non-Hermitian three layer structures and the origin of the high achievable $g$ values, we further investigate the scattering properties of our systems. Scattering processes are usually characterized by the properties (eigenvalues, eigenvectors, poles, zeros etc.) of the scattering matrix, $S$, which describes the relation between incoming and outgoing waves [35,47-51]. In our scattering (reflection/transmission) system, depending on the arrangement of the input and output ports (RCP or LCP waves), we can build our scattering matrix formalism in several ways. However, there are two cases/arrangements, which characterize two different physical processes. The first case, called here $S^{(1)}$, is related with the position of exceptional points, at which the system passes from $PT$- symmetric phase (associated with real spectrum and *PT*-symmetric eigenfunctions) to broken-*PT* phase [35] (complex spectrum, non-*PT*-symmetric eigenfunctions); i.e., the points at which the scattering matrix eigenvalues cease to be unimodular are the exceptional points, marking the change of the *PT*-phase. In the second scattering matrix case, called here $S^{(2)}$, the points at which the scattering matrix eigenvalues cease to be unimodular (called here crossing points) correspond to anisotropic transmission resonances (ATRs), in which the reflection from one side (left or right) vanishes and the transmission is unity [35], leading to zero absorption, a condition favoring dissymmetry factor enhancement. This, second *S*-matrix configuration has already been discussed quite extensively in the case of non-chiral *PT*-symmetric systems [18,42] (2x2 scattering matrix). In the case of systems involving chiral media, because of the two possible input/circular polarizations at each of the two sides of the structure, the system needs to be described by a 4x4 scattering matrix, $S^{(2)}$ [48-50]. Here $S^{(2)}$ is defined by

$$\begin{pmatrix} c_+ \\ b_- \\ c_- \\ b_+ \end{pmatrix} = S^{(2)} \begin{pmatrix} a_+ \\ d_- \\ a_- \\ d_+ \end{pmatrix} \equiv \begin{pmatrix} t_{++}^{(L)} & r_{+-}^{(R)} & 0 & 0 \\ r_{-+}^{(L)} & t_{--}^{(R)} & 0 & 0 \\ 0 & 0 & t_{--}^{(L)} & r_{-+}^{(R)} \\ 0 & 0 & r_{+-}^{(L)} & t_{++}^{(R)} \end{pmatrix} \begin{pmatrix} a_+ \\ d_- \\ a_- \\ d_+ \end{pmatrix}, \quad (5).$$

where the input and output wave amplitudes are shown in Fig. 1. The calculation of the eigenvalues of $S^{(2)}$ yields two degenerate pairs of eigenvalues, which are given by

$$\sigma_{1,2} = \tfrac{1}{2}\Big(t_{++} + t_{--} \pm \sqrt{(t_{--} - t_{++})^2 + 4r^{(L)}r^{(R)}}\Big). \quad (6)$$

Figures 5 (b) and (e) show the eigenvalues of $S^{(2)}$ for our "*PT*-symmetric" case (system of Fig. 4(b)). Although these eigenvalues depend on the chiral layer chirality, here, due to the small thickness of the chiral layer (10 nm) and the very weak chiral response (of the order of $10^{-5}$), the chiral dependence can be assumed as a small perturbation; so the eigenvalues are practically independent of chirality. If the chirality parameter is strong, i.e. in chiral metamaterials [56], the non-Hermitian/*PT*-symmetric properties, like exceptional points, ATRs etc., are strongly dependent and controllable by the chirality [49]. As can be seen in Fig. 5, at the transition points from unimodular to non-unimodular eigenvalues (multiple points in this case), which coincide with the ATRs [35], the reflection from one system side is zero while both transmittances approach unity ($T_{++} = |t_{++}|^2 \cong 1$ and $T_{--} = |t_{--}|^2 \cong 1$) – see Figs. 5 (c) & (f). Taking into account the above properties, the total absorption goes to zero, $A_+ + A_- = 2 - T_{++} - T_{--} \cong 0$; this quantity is displayed in the denominator of the dissymmetry factor $g$ [Eq. (1)]; hence the points where a $g$-peak occurs in our "PT-symmetric" system coincide with the ATR points of the system, as shown comparing Figs. 5 (a) & (d) with Figs. 5 (b) & (e), respectively. These figures clearly demonstrate and highlight the importance of the ATR points of *PT*-symmetric systems for $g$-factor maximization. Note that the results shown in Fig. 5 concern only a relatively narrow frequency region; analogous results for a much broader frequency range are shown in Fig. 8 of Appendix D. Those results demonstrate a multitude of frequency ranges where $g$-enhancement is achievable, both above and below the system exceptional point – for the *PT*-like case.

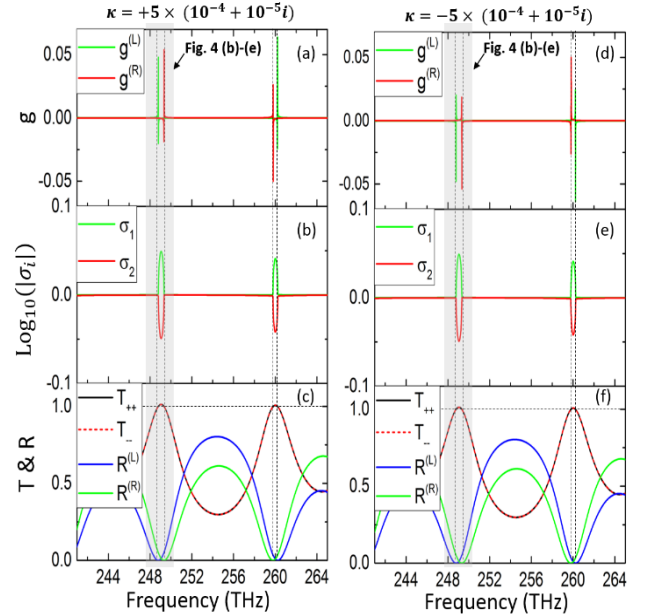

Figure 5. Panels (a) and (d) show dissymmetry factor calculations for the system of Fig. 4(b) and 4(e), with positive (left column) and negative (right column) chirality parameter. Panels (b) and (e) show the eigenvalues of the scattering matrix $S^{(2)}$ and panels (c) and (f) show the transmission and reflection power coefficients for circularly polarized plane waves impinging in both sides of the system. The vertical dashed lines correspond to the ATRs of the system.

## VI. FIELD ANALYSIS AND DISCUSSION

As was mentioned also in the introduction and in the previous section, the interplay between loss and gain media (resulting to positive and negative absorption, respectively) allows vanishing of the total absorption without reduction of the field intensity in the system. Thus, in principle, one has the potential to enhance the dissymmetry factor, enhancing thus the enantio-selectivity of the chiral light-matter interaction, keeping also high field intensity values, i.e. keeping or enhancing the overall efficiency of the interaction. To demonstrate this potential for our systems, we evaluate and plot the electric field intensity and the field chirality, *C*, along the system for the systems shown in Figs. 2 and 4 at the frequencies of the evaluated $g$-peaks shown in Fig. 4. For completeness and comparison, we present the same quantities also at CD peaks. The results are shown in Fig. 6.

More specifically, in Figs. 6 (a) and (c) we plot the normalized electric field along the *"PT*-symmetric" tri-layer structure, as well as the optical chirality (normalized by the vacuum optical chirality, $C_0$ – note that the field is everywhere in our system circularly polarized) for RCP wave propagation, at frequency 248.79 THz for waves incident from the left side (Fig. 6(a)) and at frequency 249.34 THz for waves incident from the right side (Fig. 6(c)). (Note that these frequencies coincide with the $g$- peaks of Figs. 4(b) and (e), and that very similar results are obtained also for LCP waves – not shown here). We observe that both the field intensity and the optical chirality are enhanced towards the center of the system (where the chiral layer is located) when the wave incidence is from the gain side while they decrease for incidence from the loss side, despite the fact that the transmission and reflection is in both cases the same. The reason behind this asymmetry in field and optical chirality between Figs. 6(a) and 6(c) is the coupling of the incident wave to a different system mode in the two cases, as observed and discussed also in Ref. [35]. This asymmetry can have significant impact in sensing schemes like the one discussed here (i.e. layer under sensing between loss and gain slabs), indicating a "wrong" and a "right" incidence side for enhanced sensing signals. Besides the above mentioned asymmetry, one can observe also form Figs. 6(a) and 6(c) that the fields and optical chirality are symmetric in the gain and loss slabs, as expected for *PT*-symmetric systems at anisotropic transmission resonance points [35]. (Note that the presence of the chiral layer breaks the exact *PT*-symmetry of the gain-loss bi-layer; due to its very small thickness, though, the chiral layer acts as a small perturbation, allowing the system to preserve the basic characteristics of the exact *PT* case.)

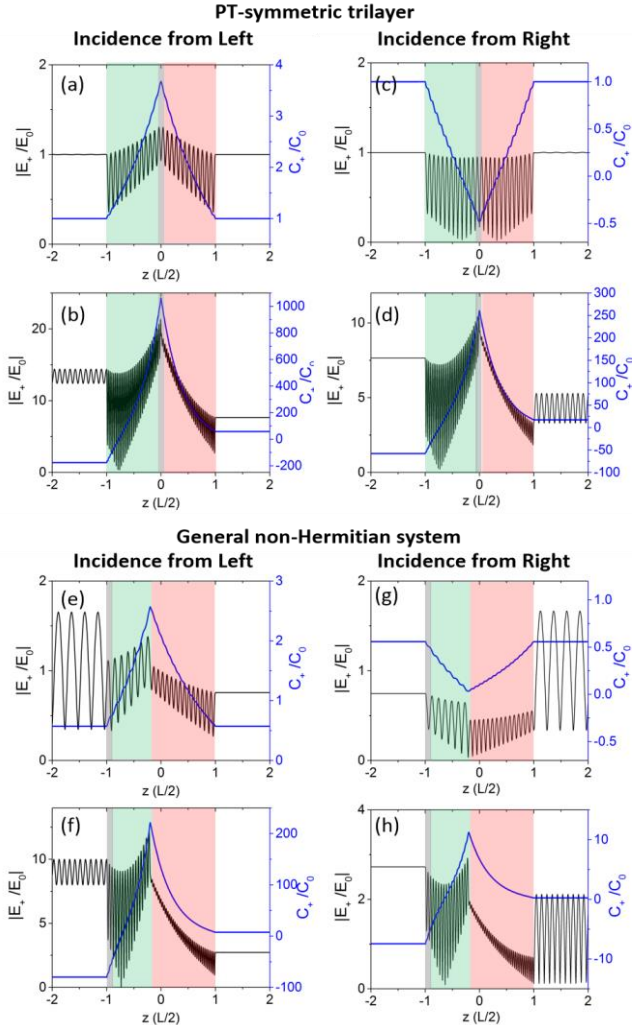

Figure 6. Normalized electric fields and optical chirality calculations for the two systems of Fig 2 (b), (c), for RCP waves incident from the left side (left column) and from the right side (right column); panels (a)-(d) correspond to the "*PT*-symmetric" system (of Fig. 2(b)) and panels (e)-(h) correspond to the general non-Hermitian system (of Fig. 2(c)). Panels (a), (c), (e), (g) concern frequencies of dissymmetry factor peaks, while panels (b), (d), (f), (h) frequencies of CD peaks. The material parameters are the same as in Fig. 2 and we assume positive sign of the chirality parameter of the thin chiral layer (both real and imaginary parts). $L$ is the total system length.

In Figs. 6 (b) and (d), we show the same results as in (a), (c) at a CD peak of the same system (at frequency 623.9 THz, where the highest CD peak is observed – see Fig. 2(b)). Here, as expected, we observe very strong field and optical chirality enhancement inside the system and in particular at its center, i.e. at the location of the chiral layer, while, again here, the incidence from the gain side leads to higher field and optical chirality values compared to the incidence from the loss side. Moreover the fields and optical chirality are not symmetric anymore between gain and loss layer, showing that this particular frequency is beyond the exceptional point of the system, as expected for *PT*-symmetric systems at the lasing threshold point and as verified also from related scattering matrix calculations (see Appendix D).

In Figs. 6 (e)-(h) we plot the normalized electric field amplitude and the optical chirality along the system for the general non-Hermitian system of Fig. 2(c) (for RCP wave propagation). Specifically, Fig. 6(e) and (g) concern *g*-peak frequencies (246.6 THz for waves incident from the left side (Fig. 6(e)) and 246.85 THz for waves incident from the right side (Fig. 6(g))), and Figs. 6 (f) and (h) a CD-resonance frequency (617 THz - see Fig. 2 (c)).

Observing the results of Figs. 6, one can see that at the *g*-peak frequencies the field intensity and the optical chirality (for both the *PT* and the non-*PT* trilayer) in the region of the chiral layer are considerably higher than their incident/vacuum values, ensuring increase of both differential and overall absorption, critical for enhanced chiro-optical response.

To investigate further the impact of the gain-loss bi-layer on the chiro-optical response of our chiral layer, we calculate the local dissymmetry factor in our systems under investigation, i.e. the dissymmetry factor evaluated by taking into account the absorption only inside the chiral layer. For that we employ the formula of Eq. (1), right-hand side (derived in Appendix F), integrating both numerator and denominator over the extent of the thin chiral layer only – see Eqs. (F10) & (F13). For simplicity, we assume only the positive values for the chirality parameter of the chiral layer; we also consider incidence from the left (gain) side, which appeared to be the "proper" side, i.e. the side maximizing the fields, according to the previous discussion (in connection with Fig. 6). For comparison, we calculate first the dissymmetry factor for the chiral layer alone – see Figs. 7(a). (The results, as expected, are the same as the corresponding ones of Fig. 4). Then, we repeat the calculations taking into account the presence of the gain and loss layers (both for the *PT*-symmetric case and for the general non-Hermitian case).

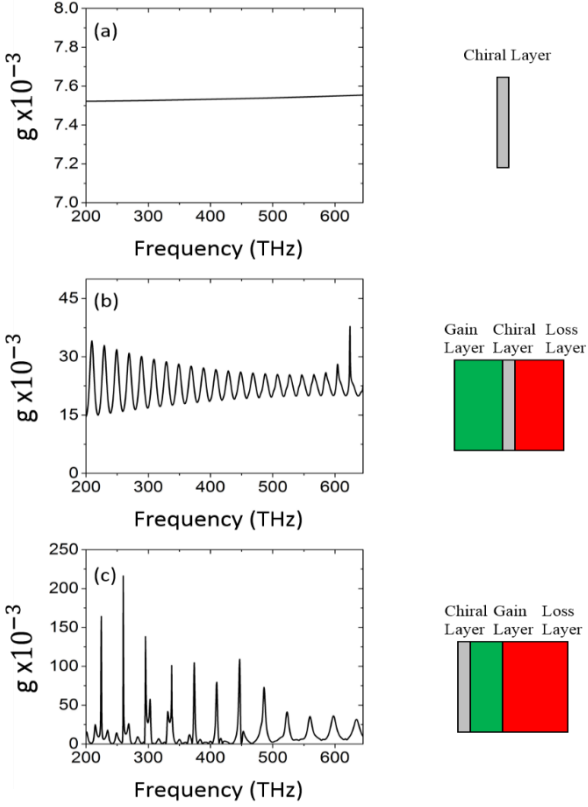

Figure 7. Kuhn's dissymmetry factor ($g$) calculated taking into account the absorption only inside the chiral layer for the systems of Fig 2. In all cases we assume positive sign of the chirality parameter (both real and imaginary parts) of the thin chiral layer.

In the case of the "*PT*-symmetric" trilayer, we observe that at a broad frequency range local $g$ shows an up to 5-fold increase due to the presence of the gain-loss bilayer, while it remains always higher than 2x the single chiral layer value. In the case of the general non-Hermitian structure (Fig. 7 (c)), the enhancement is more irregular but it achieves even higher values. Note that the $g$-enhancement shown in Figs. 4 and 7 is achieved without imposing any nanostructuring in the systems, which would lead to even more enhanced local fields and possibly enhanced local field chirality. Combining our approach (i.e. the gain approach) with such a nanostructuring (as, e.g., in [8,9]) we expect an additive effect, leading to particularly high $g$ factors and opening novel prospects in the chirality detection and synthesis.

As a closing remark, we want to mention that the purpose of the current work is mostly to illustrate the potential and perspective of the combined gain-loss systems for CD and $g$ enhancement, rather than to provide detailed guidelines for design of realistic systems. That's why we employ systems of constant gain and loss values within the whole optical range. For the realization of practical gain-loss systems for chirality detection enhancement one can use a variety of active media, depending on the desired operation frequency. Critical requirements are (a) the spectral overlapping of the gain material bandwidth with resonances of the structure, (b) a spatial (modal) coupling of gain, loss and chiral materials via close proximity, and (c) the chiral material to be placed in contact with the gain material and if possible illuminated by the gain side. To satisfy the above requirements ensuring also maximum structure performance (e.g. operation close to the lasing threshold for maximum CD enhancement) is not straightforward, requiring also guiding by detailed numerical simulations. As our results suggest though, even if the maximum performance regarding the CD is not achieved, there will be always substantial dissymmetry factor enhancement, as such an enhancement does not require operation close to the lasing threshold and can be encountered in a broad frequency range.

Regarding the gain materials that can be employed for the realization of suitable gain-loss systems, possible candidates can be dye molecules (they demonstrate gain parameter of $\gamma = 4\pi\text{Im}[n]/\lambda = 1590\ cm^{-1}$ at wavelength $\lambda = 710\ nm$ [61] and the half-width of the resonance $6000\ nm$), quantum wells or dots (with gain parameter of $\gamma = 5000\ cm^{-1}$ at wavelength $\lambda = 710\ nm$ [62]), 2D transition metal dichalcogenides (with gain parameter up to $\gamma = 3x10^4\ cm^{-1}$ at wavelength $\lambda = 640\ nm$ [63]) and others [59,64].

## VII.   CONCLUSION

We investigated non-Hermitian (i.e. combined gain-loss) systems, both "*PT*-symmetric" and non-"*PT*-symmetric", for enhancement of chiro-optical signals of a thin chiral layer incorporated in those systems. By both analytical calculations and numerical modeling, we found that in the *PT*-symmetric case the circular dichroism (CD) and the dissymmetry factor $g$ of the thin chiral layer placed between the gain and loss media can be enhanced by 75 and 7 times respectively, compared to the chiral layer alone. The largest CD enhancement was obtained at the lasing threshold of the total system, while the largest $g$-factor at its anisotropic transmission resonances; a particularly interesting and useful feature is that the $g$-factor enhancement here is associated with also total field enhancement. Departing from the exact *PT*-symmetry condition for the gain and loss layers and going to systems more amenable to practical realization, we still found considerable CD and $g$ enhancement, in some cases even more intense than in the *PT*-case. Although our approach concerns a resonant, essentially double-layer, gain-loss structure, it clearly reveals the rich possibilities offered by going to more complex structures, e.g. multi-stack *PT*-symmetric structures [35,47]; adding more layers, we may be able to create an even more enhanced response, enhancing the field and its confinement around the chiral layer.

Our results clearly demonstrate the potential of gain materials and of combined gain-loss media for chirality sensing. Moreover, our approach can be applied in combination with other optical approaches that have been proposed for chiro-optical signal enhancement, expanding the potential of those approaches and leading to novel photonic schemes for chiral bio-sensing.

Although our work concerns a simple route for chirality enhancement in a thin chiral layer, it clearly reveals the rich possibilities offered by combining gain and loss. We expect those possibilities to be even more enhanced and exploitable in other weak electromagnetic phenomena, in particular associated with non-reciprocal materials, e.g. magneto-optical materials [65,66], where the combination of gain and loss with the system resonances may lead to even larger enhancements, opening new directions in the field of non-Hermitian photonics.

This work was supported by the Hellenic Foundation for Research and Innovation (HFRI) and the General Secretariat for Research and Technology (GSRT), under the HFRI PhD Fellowship grant (GA. no. 4820), as well as by the EU-Horizon2020 FET project Ultrachiral (grant no. FETOPEN-737071) and Visorsurf. Useful communication with Prof. Alberto G. Curto is also acknowledged.

## APPENDIX A: ANALYTICAL CALCULATION OF THE REFLECTION AND TRANSMISSION FOR THE THREE-LAYER SYSTEM - THE CHIRAL TRANSFER MATRIX METHOD

To find the reflection and transmission amplitudes of the three-layer (gain, chiral, loss) structure shown in Fig. 1 of the main text, we consider waves arriving at normal incidence from either side of the structure and we solve Maxwell's equations, applying the boundary conditions at each interface. The waves propagate along *z*-direction, as illustrated in Fig. 1, and their polarization can be either linear or circular. We start with layers with arbitrary material parameters, $\varepsilon_i, \mu_i, \kappa_i$, to obtain general expressions (the subscript $i = \{G, C, L\}$ denotes the gain, chiral and loss regions respectively). Due to the two possible input circular polarizations (denoted as RCP/+ and LCP/-) at each of the two sides of the system, the full description of the scattering process requires 4 input and 4 output ports; hence the system is described by a 4x4 scattering matrix, consisting of 8 transmission (*t*) and 8 reflection (*r*) coefficients.

In order to determine the reflection and transmission coefficients, we assume that we have either RCP/+ or LCP/- polarized incident plane waves, arriving from the left side (see Fig1):

$$\begin{cases} \boldsymbol{E}_{in}^{air} = a_\pm \hat{\boldsymbol{e}}_\pm e^{ikz} \\ \boldsymbol{H}_{in}^{air} = \mp i Z^{-1} a_\pm \hat{\boldsymbol{e}}_\pm e^{ikz} \end{cases} \quad (A1)$$

(the structure is embedded in air and the subscript (in) denotes the incident wave). The reflected electromagnetic fields can be expressed as:

$$\begin{cases} \boldsymbol{E}_{ref}^{air} = -b_+ \hat{\boldsymbol{e}}_- e^{-ikz} - b_- \hat{\boldsymbol{e}}_+ e^{-ikz} \\ \boldsymbol{H}_{ref}^{air} = -iZ^{-1}(-b_+)\hat{\boldsymbol{e}}_- e^{-ikz} + iZ^{-1}(-b_-)\hat{\boldsymbol{e}}_+ e^{-ikz} \end{cases} \quad (A2)$$

where $a_\pm, b_\pm$ are the amplitudes of the ingoing and outgoing, respectively, RCP/+ and LCP/- waves (as observed from the source point) and $\hat{\boldsymbol{e}}_\pm = (\hat{\boldsymbol{x}} \pm i\hat{\boldsymbol{y}})/\sqrt{2}$. A similar expression can be employed for electric and magnetic fields inside the gain, the loss and the chiral layers, where the total field is a sum of a forward propagating and a backward propagating wave. In the chiral layer in particular, due to two the different refractive indices $n_\pm = \sqrt{\mu\varepsilon} \pm \kappa$, we have different wavevectors for the RCP and LCP components. Thus in the chiral layer the electromagnetic field can be expressed as:

$$\begin{cases} \boldsymbol{E}_{forw}^{chiral} = E_+^{(L)} \hat{\boldsymbol{e}}_+ e^{ik_+ z} + E_- \hat{\boldsymbol{e}}_- e^{ik_- z} \\ \boldsymbol{H}_{forw}^{chiral} = \mp i Z^{-1}(E_+^{(L)} \hat{\boldsymbol{e}}_+ e^{ik_+ z} + E_- \hat{\boldsymbol{e}}_- e^{ik_- z}) \end{cases} \quad (A3)$$

and

$$\begin{cases} \boldsymbol{E}_{back}^{chiral} = -E_+^{(R)} \hat{\boldsymbol{e}}_- e^{-ik_- z} - E_- \hat{\boldsymbol{e}}_+ e^{-ik_+ z} \\ \boldsymbol{H}_{back}^{chiral} = \mp i Z^{-1}(-E_+^{(R)} \hat{\boldsymbol{e}}_- e^{-ik_- z} - E_- \hat{\boldsymbol{e}}_+ e^{-ik_+ z}) \end{cases} \quad (A4).$$

The transmitted electromagnetic fields can be expressed as:

$$\begin{cases} \boldsymbol{E}_{tr}^{air} = c_+ \hat{\boldsymbol{e}}_+ e^{ikz} + c_- \hat{\boldsymbol{e}}_- e^{ikz} \\ \boldsymbol{H}_{tr}^{air} = -iZ^{-1} c_+ \hat{\boldsymbol{e}}_+ e^{ikz} + iZ^{-1} c_- \hat{\boldsymbol{e}}_- e^{ikz} \end{cases} \quad (A5).$$

where $c_\pm$ are the amplitudes of the transmitted waves (see Fig. 1). Equating the tangential components of ***E*** and ***H*** ($E_x, E_y, H_x, H_y$) at the four structure interfaces, we obtain a 16x16 linear system of equations with 72 out of 256 non-zero elements, which is solved analytically or numerical (using for example Gaussian-elimination with partial pivot method), giving the scattering coefficients (for more detailed analysis regarding the methodology see [49] and [9]). The reflection and transmission coefficients when the incident wave arrives from the right side of the structure can be obtained by applying the same analysis as above or, more easily, by exchanging the material parameters in the gain-loss layers.

Although in this work we focus on the analytical derivation (although cumbersome yet straightforward) of the reflection and transmission coefficients for a three-layer system, we consider useful to provide a derivation approach that can be extended to multilayer systems, being also more compact in the formalism. Such an approach is provided by the transfer matrix method. Therefore, we present below the transfer matrix formalism in circular polarization basis, suitable for the analysis of electromagnetic wave propagation in multilayer structures (including chiral layers), under normal incidence. The transfer matrix *M* relates the fields on one side of an interface to those of the other side (see Fig. 1 for the definition of field amplitudes). Because the electric field in the circular polarization basis is described by four waves, two (LCP/+ and RCP/- as a source view) incident on the interface and the other two outgoing, the matching matrix *M* is a 4x4 matrix, of the form

$$\begin{bmatrix} a_+ \\ a_- \\ b_+ \\ b_- \end{bmatrix}_i = M_{i,i+1} \begin{bmatrix} c_+ \\ c_- \\ d_+ \\ d_- \end{bmatrix}_{i+1} \quad (A7)$$

where $i$ (counting the layers) denotes the layer left-to the interface, and $M_{i,i+1}$ is block symmetric of the form $M_{i,i+1} = \begin{bmatrix} A & B \\ B & A \end{bmatrix}$, with the corresponding submatrices given by

$$A = \begin{bmatrix} \frac{1}{2}\left(\frac{Z_i}{Z_{i+1}}+1\right) & 0 \\ 0 & \frac{1}{2}\left(\frac{Z_i}{Z_{i+1}}+1\right) \end{bmatrix} \quad (A8)$$

$$B = \begin{bmatrix} 0 & \frac{1}{2}\left(\frac{Z_i}{Z_{i+1}}-1\right) \\ \frac{1}{2}\left(\frac{Z_i}{Z_{i+1}}-1\right) & 0 \end{bmatrix}. \quad (A9)$$

Referring to Fig. 1, the fields on the left side of the system and those on the right side are connected with the total transfer matrix, $M^{(total)}$. For the construction of $M^{(total)}$, besides the transfer matrices of the individual interfaces, one has to consider also the phase matrices. A phase matrix $P_i$ accounts for the phase contributed by forward- and backward-waves propagating inside the medium $i$, with a thickness of $d_i$, and is given by

$$P_i = \begin{bmatrix} e^{ik_+^{(i)}d_i} & 0 & 0 & 0 \\ 0 & e^{ik_-^{(i)}d_i} & 0 & 0 \\ 0 & 0 & e^{-ik_+^{(i)}d_i} & 0 \\ 0 & 0 & 0 & e^{-ik_-^{(i)}d_i} \end{bmatrix}. \quad (A10)$$

For non-chiral layers the $P_i$ matrices can be simplified by taking $k_+^{(i)} = k_-^{(i)} = k^{(i)}$. The system of multilayers can then be solved by the ordered multiplication of interface transfer matrices $M_{i,i+1}$ and the in-layer phase matrices $P_i$. The resulting transfer matrix $M^{(total)}$ contains all the wave reflection and transmission information for the system. For the gain-chiral-loss system of Fig. 1 the total transfer matrix is given by

$$\begin{bmatrix} a_+ \\ a_- \\ b_+ \\ b_- \end{bmatrix} = M^{(total)} \begin{bmatrix} c_+ \\ c_- \\ d_+ \\ d_- \end{bmatrix} \equiv M_{01} P_1 M_{12} P_2 M_{23} P_3 M_{34} \begin{bmatrix} c_+ \\ c_- \\ d_+ \\ d_- \end{bmatrix}. \quad (A11)$$

where the subscripts 0 and 4 correspond to the layers before and after the structure, respectively (both air in our case). Having evaluated the transfer matrix $M^{(total)}$ one can decompose it in four 2x2 submatrices, of the form $M^{(total)} = \begin{bmatrix} T_1 & R_1 \\ R_2 & T_2 \end{bmatrix}$. Then the reflection and transmission matrices for incidence from the left side of the system are given by $R^{(L)} = \begin{bmatrix} r_{++}^{(L)} & r_{+-}^{(L)} \\ r_{-+}^{(L)} & r_{--}^{(L)} \end{bmatrix} = R_2(T_1)^{-1}$ and $T^{(L)} = \begin{bmatrix} t_{++}^{(L)} & t_{+-}^{(L)} \\ t_{-+}^{(L)} & t_{--}^{(L)} \end{bmatrix} = (T_1)^{-1}$ while for incidence from the right side are given by $R^{(R)} = \begin{bmatrix} r_{++}^{(R)} & r_{+-}^{(R)} \\ r_{-+}^{(R)} & r_{--}^{(R)} \end{bmatrix} = -(T_1)^{-1}R_1$ and $T^{(R)} = \begin{bmatrix} t_{++}^{(R)} & t_{+-}^{(R)} \\ t_{-+}^{(R)} & t_{--}^{(R)} \end{bmatrix} = T_2 - R_2(T_1)^{-1}R_1$ [67,68].

For layers of isotropic chiral materials (reciprocal) and normal incidence, it is in general valid, that $t_{++}^{(L)} = t_{++}^{(R)}$, $t_{--}^{(L)} = t_{--}^{(R)}$ and $r_{+-}^{(L)} = r_{-+}^{(L)}$, $r_{+-}^{(R)} = r_{-+}^{(R)}$ while the remaining eight coefficients $r_{++}^{(L)} = r_{--}^{(L)} = r_{++}^{(R)} = r_{--}^{(R)} = t_{+-}^{(L)} = t_{-+}^{(L)} = t_{+-}^{(R)} = t_{-+}^{(R)} = 0$ regardless of the side of incidence. The analytical expressions of $r$, $t$ for circularly and linearly polarized waves are listed below.

### Scattering coefficients for circularly polarized waves

**Incident from left**

$$t_{++}^{(L)} = -\frac{16\, e^{i[d_C(k-k_G-k_L)+d(2k-k_G-k_L)]}}{Z_G Z_C Z_L (A_1+A_2+B_1+B_2)} e^{ik_{C_1}d_C}$$
$$t_{--}^{(L)} = -\frac{16\, e^{i[d_C(k-k_G-k_L)+d(2k-k_G-k_L)]}}{Z_G Z_C Z_L (A_1+A_2+B_1+B_2)} e^{ik_{C_2}d_C}$$
$$t_{-+}^{(L)} = t_{+-}^{(L)} = 0 \quad (A12)$$
$$r_{+-}^{(L)} = r_{-+}^{(L)} = \frac{e^{-2ik\left(d+\frac{d_C}{2}\right)}\left(C_1^{(L)}+C_2^{(L)}+D_1^{(L)}+D_2^{(L)}\right)}{(A_1+A_2+B_1+B_2)}$$
$$r_{++}^{(L)} = r_{--}^{(L)} = 0$$

**Incident from right**

$$t_{++}^{(R)} = -\frac{16\, e^{i[d_C(k-k_G-k_L)+d(2k-k_G-k_L)]}}{Z_G Z_C Z_L (A_1+A_2+B_1+B_2)} e^{ik_{C_1}d_C}$$
$$t_{--}^{(R)} = -\frac{16\, e^{i[d_C(k-k_G-k_L)+d(2k-k_G-k_L)]}}{Z_G Z_C Z_L (A_1+A_2+B_1+B_2)} e^{ik_{C_2}d_C}$$
$$t_{-+}^{(R)} = t_{+-}^{(R)} = 0 \quad (A13)$$
$$r_{+-}^{(R)} = r_{-+}^{(R)} = \frac{e^{-2ik\left(d+\frac{d_C}{2}\right)}\left(C_1^{(R)}+C_2^{(R)}+D_1^{(R)}+D_2^{(R)}\right)}{(A_1+A_2+B_1+B_2)}$$
$$r_{++}^{(R)} = r_{--}^{(R)} = 0$$

### Scattering coefficients for linearly polarized waves

**Incident from left, x-polarized**

**Incident from right, x-polarized**

$$t_{xx}^{(L)} = -\frac{8\,e^{-i[d_C(k-k_G-k_L)+d(2k-k_G-k_L)]}}{Z_G Z_C Z_L (A_1+A_2+B_1+B_2)}(e^{ik_{C1}d_C}+e^{ik_{C2}d_C})$$

$$t_{yx}^{(L)} = -i\frac{8\,e^{-i[d_C(k-k_G-k_L)+d(2k-k_G-k_L)]}}{Z_G Z_C Z_L (A_1+A_2+B_1+B_2)}(e^{ik_{C1}d_C}-e^{ik_{C2}d_C}) \quad \text{(A14)}$$

$$r_{xx}^{(L)} = \frac{e^{-2ik\left(d+\frac{d_C}{2}\right)}\left(C_1^{(L)}+C_2^{(L)}+D_1^{(L)}+D_2^{(L)}\right)}{(A_1+A_2+B_1+B_2)}$$

$$r_{yx}^{(L)} = 0$$

$$t_{xx}^{(R)} = t_{xx}^{(L)}$$
$$t_{yx}^{(R)} = -t_{yx}^{(L)}$$
$$r_{xx}^{(R)} = \frac{e^{-2ik\left(d+\frac{d_C}{2}\right)}\left(C_1^{(R)}+C_2^{(R)}+D_1^{(R)}+D_2^{(R)}\right)}{(A_1+A_2+B_1+B_2)} \quad \text{(A15)}$$
$$r_{yx}^{(R)} = 0$$

**Incident from left, y-polarized**

$$\begin{aligned} t_{yy}^{(L)} &= t_{xx}^{(L)} \\ t_{xy}^{(L)} &= -t_{yx}^{(L)} \\ r_{yy}^{(L)} &= r_{xx}^{(L)} \\ r_{xy}^{(L)} &= r_{yx}^{(L)} \end{aligned} \quad \text{(A16)}$$

**Incident from right, y-polarized**

$$\begin{aligned} t_{yy}^{(R)} &= t_{xx}^{(R)} \\ t_{xy}^{(R)} &= -t_{yx}^{(R)} \\ r_{yy}^{(R)} &= r_{xx}^{(R)} \\ r_{xy}^{(R)} &= r_{yx}^{(R)} \end{aligned} \quad \text{(A17)}$$

In the above formulas, (A12)-(A17), the superscripts (*L*) and (*R*) denote incidence from the left and right side of the system, respectively, the first subscript in *r, t* denotes the output polarization and the second the input polarization, and

$$A_1 = e^{2i\left(d+\frac{d_C}{2}\right)(k_G+k_L)}\left(\frac{1}{Z_G}-1\right)\left(\frac{1}{Z_G}-\frac{1}{Z_C}\right)\left(\frac{1}{Z_C}-\frac{1}{Z_L}\right)\left(\frac{1}{Z_L}-1\right) - e^{2i[k_L d+(k_G+k_L)d_C/2]}\left(\frac{1}{Z_G}+1\right)\left(\frac{1}{Z_G}+\frac{1}{Z_C}\right)\left(\frac{1}{Z_C}-\frac{1}{Z_L}\right)\left(\frac{1}{Z_L}-1\right)$$

$$A_2 = -e^{id_C(k_G+k_{C1}+k_{C2}+k_L)}\left(\frac{1}{Z_G}+1\right)\left(\frac{1}{Z_G}-\frac{1}{Z_C}\right)\left(\frac{1}{Z_C}-\frac{1}{Z_L}\right)\left(\frac{1}{Z_L}+1\right) + e^{2i\left[k_G d+\frac{(k_G+k_{C1}+k_{C2}+k_L)d_C}{2}\right]}\left(\frac{1}{Z_G}-1\right)\left(\frac{1}{Z_G}+\frac{1}{Z_C}\right)\left(\frac{1}{Z_C}-\frac{1}{Z_L}\right)\left(\frac{1}{Z_L}+1\right)$$

$$B_1 = -e^{2i\left[k_L d+\frac{(k_G+k_{C1}+k_{C2}+k_L)d_C}{2}\right]}\left(\frac{1}{Z_G}+1\right)\left(\frac{1}{Z_G}-\frac{1}{Z_C}\right)\left(\frac{1}{Z_C}+\frac{1}{Z_L}\right)\left(\frac{1}{Z_L}-1\right) + e^{2i\left[(k_G+k_L)d+\frac{(k_G+k_{C1}+k_{C2}+k_L)d_C}{2}\right]}\left(\frac{1}{Z_G}-1\right)\left(\frac{1}{Z_G}+\frac{1}{Z_C}\right)\left(\frac{1}{Z_C}+\frac{1}{Z_L}\right)\left(\frac{1}{Z_L}-1\right) \quad \text{(A18)}$$

$$B_2 = e^{2i[k_G d+(k_G+k_L)d_C/2]}\left(\frac{1}{Z_G}-1\right)\left(\frac{1}{Z_G}-\frac{1}{Z_C}\right)\left(\frac{1}{Z_C}+\frac{1}{Z_L}\right)\left(\frac{1}{Z_L}+1\right) - e^{i(k_G+k_L)d_C}\left(\frac{1}{Z_G}+1\right)\left(\frac{1}{Z_G}+\frac{1}{Z_C}\right)\left(\frac{1}{Z_C}+\frac{1}{Z_L}\right)\left(\frac{1}{Z_L}+1\right)$$

$$C_1^{(L)} = -e^{2i\left(d+\frac{d_C}{2}\right)(k_G+k_L)}\left(\frac{1}{Z_G}+1\right)\left(\frac{1}{Z_G}-\frac{1}{Z_C}\right)\left(\frac{1}{Z_C}-\frac{1}{Z_L}\right)\left(\frac{1}{Z_L}-1\right) + e^{2i[k_L d+(k_G+k_L)d_C/2]}\left(\frac{1}{Z_G}-1\right)\left(\frac{1}{Z_G}+\frac{1}{Z_C}\right)\left(\frac{1}{Z_C}-\frac{1}{Z_L}\right)\left(\frac{1}{Z_L}-1\right)$$

$$C_2^{(L)} = e^{id_C(k_G+k_{C1}+k_{C2}+k_L)}\left(\frac{1}{Z_G}-1\right)\left(\frac{1}{Z_G}-\frac{1}{Z_C}\right)\left(\frac{1}{Z_C}-\frac{1}{Z_L}\right)\left(\frac{1}{Z_L}+1\right) - e^{2i[k_G d+(k_G+k_{C1}+k_{C2}+k_L)d_C/2]}\left(\frac{1}{Z_G}+1\right)\left(\frac{1}{Z_G}+\frac{1}{Z_C}\right)\left(\frac{1}{Z_C}-\frac{1}{Z_L}\right)\left(\frac{1}{Z_L}+1\right)$$

$$D_1^{(L)} = e^{2i\left[k_L d+\frac{(k_G+k_{C1}+k_{C2}+k_L)d_C}{2}\right]}\left(\frac{1}{Z_G}-1\right)\left(\frac{1}{Z_G}-\frac{1}{Z_C}\right)\left(\frac{1}{Z_C}+\frac{1}{Z_L}\right)\left(\frac{1}{Z_L}-1\right) - e^{2i\left[(k_G+k_L)d+\frac{(k_G+k_{C1}+k_{C2}+k_L)d_C}{2}\right]}\left(\frac{1}{Z_G}+1\right)\left(\frac{1}{Z_G}+\frac{1}{Z_C}\right)\left(\frac{1}{Z_C}+\frac{1}{Z_L}\right)\left(\frac{1}{Z_L}-1\right) \quad \text{(A19)}$$

$$D_2^{(L)} = -e^{2i\left[k_G d+\frac{(k_G+k_L)d_C}{2}\right]}\left(\frac{1}{Z_G}+1\right)\left(\frac{1}{Z_G}-\frac{1}{Z_C}\right)\left(\frac{1}{Z_C}+\frac{1}{Z_L}\right)\left(\frac{1}{Z_L}+1\right) + e^{i(k_G+k_L)d_C}\left(\frac{1}{Z_G}-1\right)\left(\frac{1}{Z_G}+\frac{1}{Z_C}\right)\left(\frac{1}{Z_C}+\frac{1}{Z_L}\right)\left(\frac{1}{Z_L}+1\right)$$

$$C_1^{(R)} = -e^{2i\left(d+\frac{d_C}{2}\right)(k_G+k_L)}\left(\frac{1}{Z_L}+1\right)\left(\frac{1}{Z_L}-\frac{1}{Z_C}\right)\left(\frac{1}{Z_C}-\frac{1}{Z_G}\right)\left(\frac{1}{Z_G}-1\right) + e^{2i[k_G d+(k_G+k_L)d_C/2]}\left(\frac{1}{Z_L}-1\right)\left(\frac{1}{Z_L}+\frac{1}{Z_C}\right)\left(\frac{1}{Z_C}-\frac{1}{Z_G}\right)\left(\frac{1}{Z_G}-1\right)$$

$$C_2^{(R)} = e^{id_C(k_G+k_{C1}+k_{C2}+k_L)}\left(\frac{1}{Z_L}-1\right)\left(\frac{1}{Z_L}-\frac{1}{Z_C}\right)\left(\frac{1}{Z_C}-\frac{1}{Z_G}\right)\left(\frac{1}{Z_G}+1\right) - e^{2i[k_L d+(k_G+k_{C1}+k_{C2}+k_L)d_C/2]}\left(\frac{1}{Z_G}+1\right)\left(\frac{1}{Z_G}+\frac{1}{Z_C}\right)\left(\frac{1}{Z_C}-\frac{1}{Z_L}\right)\left(\frac{1}{Z_L}+1\right)$$

$$D_1^{(R)} = e^{2i\left[k_G d+\frac{(k_G+k_{C1}+k_{C2}+k_L)d_C}{2}\right]}\left(\frac{1}{Z_L}-1\right)\left(\frac{1}{Z_L}-\frac{1}{Z_C}\right)\left(\frac{1}{Z_C}+\frac{1}{Z_G}\right)\left(\frac{1}{Z_G}-1\right) - e^{2i\left[(k_G+k_L)d+\frac{(k_G+k_{C1}+k_{C2}+k_L)d_C}{2}\right]}\left(\frac{1}{Z_L}+1\right)\left(\frac{1}{Z_L}+\frac{1}{Z_C}\right)\left(\frac{1}{Z_C}+\frac{1}{Z_G}\right)\left(\frac{1}{Z_G}-1\right) \quad \text{(A20)}$$

$$D_2^{(R)} = -e^{2i\left[k_L d+\frac{(k_G+k_L)d_C}{2}\right]}\left(\frac{1}{Z_L}+1\right)\left(\frac{1}{Z_L}-\frac{1}{Z_C}\right)\left(\frac{1}{Z_C}+\frac{1}{Z_G}\right)\left(\frac{1}{Z_G}+1\right) + e^{i(k_G+k_L)d_C}\left(\frac{1}{Z_L}-1\right)\left(\frac{1}{Z_L}+\frac{1}{Z_C}\right)\left(\frac{1}{Z_C}+\frac{1}{Z_G}\right)\left(\frac{1}{Z_G}+1\right)$$

In the above relations $Z_i = \sqrt{\mu_0\mu_i/\varepsilon_0\varepsilon_i}$, $i = \{G,C,L\}$, is the wave impedance and $k_G = \frac{\omega(\sqrt{\varepsilon_G\mu_G})}{c}$, $k_L = \frac{\omega(\sqrt{\varepsilon_L\mu_L})}{c}$ and $k_{C1} = \frac{\omega(\sqrt{\varepsilon_C\mu_C}+\kappa)}{c}$, $k_{C2} = \frac{\omega(\sqrt{\varepsilon_C\mu_C}-\kappa)}{c}$ are the wave vectors in the gain, loss and chiral regions, respectively.

**APPENDIX B: DERIVATION OF EQs. (2) & (3)-(4)**

In this section we calculate the circular dichroism, CD, and the dissymmetry factor, *g*, for the system shown in Fig. 1 of the main text with *PT*-symmetric gain-loss layers. In order to calculate the CD and *g* it is necessary to calculate the absorption amplitudes for RCP and LCP waves. According to the scattering matrix, *S*, theory [67], the main diagonal of the difference $\hat{1} - S^\dagger S$ (where $\hat{1}$ is the identity matrix and † means transpose and complex conjugate) contains the absorption amplitudes for different circular polarizations (RCP/+)-(LCP/-) and sides of incidence. We express them as

$$\begin{aligned} A_+^{(L)} &= 1 - \left|r_{-+}^{(L)}\right|^2 - \left|t_{++}^{(L)}\right|^2 \\ A_-^{(R)} &= 1 - \left|r_{+-}^{(R)}\right|^2 - \left|t_{--}^{(R)}\right|^2 \\ A_-^{(L)} &= 1 - \left|r_{+-}^{(L)}\right|^2 - \left|t_{--}^{(L)}\right|^2 \\ A_+^{(R)} &= 1 - \left|r_{-+}^{(R)}\right|^2 - \left|t_{++}^{(R)}\right|^2 \end{aligned} \quad \text{(B1)}$$

In the previous section, we found that $t_{++}^{(L)} = t_{++}^{(R)} \equiv t_{++}$ and $t_{--}^{(L)} = t_{--}^{(R)} \equiv t_{--}$ (reciprocal system) while $r_{-+}^{(L)} = r_{+-}^{(L)} \equiv$

$r^{(L)}$ and $r_{+-}^{(R)} = r_{-+}^{(R)} \equiv r^{(R)}$; therefore the $CD = A_+ - A_-$ becomes

$$CD^{(L)} = CD^{(R)} \equiv CD = |t_{--}|^2 - |t_{++}|^2 \quad (B2)$$

$$CD = \frac{256 e^{-2[\text{Im}(k_{C1}d_C) + \text{Im}(k_{C2}d_C) + \text{Im}(d_C(k-k_G-k_L) + d(2k-k_G-k_L))]}}{|Z_G Z_C Z_L (A_1 + A_2 + B_1 + B_2)|^2} \left[ e^{2\text{Im}(k_{C1}d_C)} - e^{2\text{Im}(k_{C2}d_C)} \right] \quad (B3)$$

In the case of *PT*-symmetric gain-loss layers the wave vectors obey $k_G = (k_L)^*$; moreover $k_{C1} = \omega(n_C + \kappa)/c$, $k_{C2} = \omega(n_C - \kappa)/c$, where $n_C$ and $\kappa$ the non-chiral refractive index and the chirality parameter of the chiral layer, respectively. After some algebra, the term $e^{2\text{Im}(k_{C1}d_C)} - e^{2\text{Im}(k_{C2}d_C)}$ in Eq. (B3) can be simplified as $2e^{2kd_C\text{Im}[n_C]}\sinh(2kd_C\text{Im}[\kappa])$ and the term $e^{-2[\text{Im}(k_{C1}d_C)+\text{Im}(k_{C2}d_C)+\text{Im}(d_C(k-k_G-k_L)+d(2k-k_G-k_L))]}$ can be expressed as $e^{-4kd_C\text{Im}[n_C]}$. Hence, the circular dichroism in the *PT* case can be written (Eq. (2) of the main text) as

$$CD = \frac{512 e^{-2kd_C\text{Im}[n_C]}}{|Z_G Z_C Z_L (A_1 + A_2 + B_1 + B_2)|^2} \sinh(2kd_C\text{Im}[\kappa]), (B4)$$

Inserting the transmission coefficients from Eq. (A6) into Eq. (B2) we find where the terms $A_1, A_2, B_1, B_2$ are given in the previous section and are totally independent of the chirality parameter, $\kappa$.

For the calculation of the dissymmetry factor, $g$, defined as [21]

$$g = \frac{2CD}{A_+ + A_-}, \quad (B5)$$

besides circular dichroism one needs to calculate also the total absorption $A_+ + A_-$ (for RCP and LCP waves) for incidence from both system sides. From Eqs. (B1) we see that the total absorption depends on the reflectance and therefore depends on the side of incidence, as

$$\begin{aligned} A_+^{(L)} + A_-^{(L)} &= 1 - \left|r_{-+}^{(L)}\right|^2 - \left|t_{++}^{(L)}\right|^2 + 1 - \left|r_{+-}^{(L)}\right|^2 - \left|t_{--}^{(L)}\right|^2 \equiv 2 - |t_{++}|^2 - |t_{--}|^2 - 2|r^{(L)}|^2 \\ A_+^{(R)} + A_-^{(R)} &= 1 - \left|r_{-+}^{(R)}\right|^2 - \left|t_{++}^{(R)}\right|^2 + 1 - \left|r_{+-}^{(R)}\right|^2 - \left|t_{--}^{(R)}\right|^2 \equiv 2 - |t_{++}|^2 - |t_{--}|^2 - 2|r^{(R)}|^2 \end{aligned} \quad (B6)$$

By substituting equations (B3) and (B6) into (B5) and after extensive but straightforward calculations, we obtain

$$g^{(L)} = -256 \left[ e^{2\text{Im}(k_{C1}d_C)} - e^{2\text{Im}(k_{C2}d_C)} \right] \left[ \frac{1}{128(e^{2\text{Im}(k_{C1}d_C)} + e^{2\text{Im}(k_{C2}d_C)})} - \frac{1}{P e^{2\text{Im}(d_C(k-k_G+k_{C1}+k_{C2}-k_L)+d(2k-k_G-k_L))}} \right.$$
$$\left. + \frac{1}{Q^{(L)} e^{2\text{Im}(d_C(k-k_G+k_{C1}+k_{C2}-k_L)+d(2k-k_G-k_L))}} \right] \quad (B7)$$

$$g^{(R)} = -256 \left[ e^{2\text{Im}(k_{C1}d_C)} - e^{2\text{Im}(k_{C2}d_C)} \right] \left[ \frac{1}{128(e^{2\text{Im}(k_{C1}d_C)} + e^{2\text{Im}(k_{C2}d_C)})} - \frac{1}{P e^{2\text{Im}(d_C(k-k_G+k_{C1}+k_{C2}-k_L)+d(2k-k_G-k_L))}} \right.$$
$$\left. + \frac{1}{Q^{(R)} e^{2\text{Im}(d_C(k-k_G+k_{C1}+k_{C2}-k_L)+d(2k-k_G-k_L))}} \right] \quad (B8)$$

where $P = |Z_G Z_C Z_L (A_1 + A_2 + B_1 + B_2)|^2$, $Q^{(L)} = \left|Z_G Z_C Z_L \left(C_1^{(L)} + C_2^{(L)} + D_1^{(L)} + D_2^{(L)}\right)\right|^2$ and $Q^{(R)} = \left|Z_G Z_C Z_L \left(C_1^{(R)} + C_2^{(R)} + D_1^{(R)} + D_2^{(R)}\right)\right|^2$. Taking into account the *PT*-symmetry of the gain-loss layers, Equations (B7) and (B8) can be simplified as

$$g^{(L)} = -2\tanh(2kd_C\text{Im}[\kappa]) + 512 e^{-2kd_C\text{Im}[n_C]}\sinh(2kd_C\text{Im}[\kappa]) \left[\frac{1}{P} + \frac{1}{Q^{(L)}}\right] \quad (B9)$$

$$g^{(R)} = -2\tanh(2kd_C\text{Im}[\kappa]) + 512 e^{-2kd_C\text{Im}[n_C]}\sinh(2kd_C\text{Im}[\kappa]) \left[\frac{1}{P} + \frac{1}{Q^{(R)}}\right] \quad (B10)$$

**APPENDIX C: SCATTERING MATRIX APPROACH**

It is often convenient to describe scattering systems in terms of their scattering matrix, $S$. Among others, $S$ can be used to identify and/or analyse exotic scattering phenomena such as *PT*-symmetric and broken-*PT*-symmetric phases, exceptional points, anisotropic transmission resonances etc. [35,50-51]. By definition the scattering matrix of a system describes the relation between its incoming and outgoing waves. In systems involving chiral media (and therefore interacting differently with RCP and LCP waves), such as the one of Fig. 1 of the main text, because of the two possible circular polarizations at each of the two sides of a system, it should be described by a 4x4 scattering matrix $S$. Depending on the arrangement of the input and output ports (RCP or LCP waves), we can build the scattering matrix formalism in several ways. However, as the main information we want to extract is related with the position of the exceptional points (where a $PT$-system passes from $PT$-symmetric phase to broken-$PT$ phase) and of the anisotropic transmission resonances, we will list here only the two relevant (as shown in the literature [35]) scattering matrix configurations along with their eigenvalues. The first configuration (Case I), denoted here by $S^{(1)}$, is related to the position of exceptional point/points; i.e. the points at which $S^{(1)}$ eigenvalues stop being unimodular are exceptional points, where two or more eigenvalues and eigenvectors coincide. In this definition, the reflection coefficients are on the diagonal, and the outgoing waves are related to the incident waves through time-reversal. This behavior mimic the 2x2 matrix representation of $PT$-symmetric Hamiltonian [69] or $PT$-symmetric coupled optical waveguides [44] with the different reflected coefficients comes due to different wave impedances in the gain-loss layers. The $S$ matrix of case II ($S^{(2)}$) gives the positions of the anisotropic transmission resonances (ATRs) of the system (i.e. its crossing points, where eigenvalues turn from unimodular to non-unimodular and vice versa are ATRs), which coincide with the peaks of the $g$-factor, as discussed in the main text. Here, the transmission coefficients are on the diagonal as a result a different criterion for $PT$-symmetric breaking phases [35]. $S^{(1)}$ and $S^{(2)}$ are defined as shown below.

$$\text{Case I:} \begin{pmatrix} b_- \\ c_+ \\ b_+ \\ c_- \end{pmatrix} = S^{(1)} \begin{pmatrix} a_+ \\ d_- \\ a_- \\ d_+ \end{pmatrix} \equiv \begin{pmatrix} r^{(L)}_{-+} & t^{(R)}_{--} & 0 & 0 \\ t^{(L)}_{++} & r^{(R)}_{+-} & 0 & 0 \\ 0 & 0 & r^{(L)}_{+-} & t^{(R)}_{++} \\ 0 & 0 & t^{(L)}_{--} & r^{(R)}_{-+} \end{pmatrix} \begin{pmatrix} a_+ \\ d_- \\ a_- \\ d_+ \end{pmatrix}, \text{(C1)}$$

with two degenerate pairs of eigenvalues, $\sigma$:

$$\sigma^{(1)}_{1,2} = \frac{1}{2}\left(r^{(L)} + r^{(R)} \pm \sqrt{(r^{(L)} - r^{(R)})^2 + 4 t_{++} t_{--}}\right), \text{(C2)}$$

$$\text{Case II:} \begin{pmatrix} c_+ \\ b_- \\ c_- \\ b_+ \end{pmatrix} = S^{(2)} \begin{pmatrix} a_+ \\ d_- \\ a_- \\ d_+ \end{pmatrix} \equiv \begin{pmatrix} t^{(L)}_{++} & r^{(R)}_{+-} & 0 & 0 \\ r^{(L)}_{-+} & t^{(R)}_{--} & 0 & 0 \\ 0 & 0 & t^{(L)}_{--} & r^{(R)}_{-+} \\ 0 & 0 & r^{(L)}_{+-} & t^{(R)}_{++} \end{pmatrix} \begin{pmatrix} a_+ \\ d_- \\ a_- \\ d_+ \end{pmatrix}, \text{(C3)}$$

with two degenerate pairs of eigenvalues:

$$\sigma^{(2)}_{1,2} = \frac{1}{2}\left(t_{++} + t_{--} \pm \sqrt{(t_{--} - t_{++})^2 + 4 r^{(L)} r^{(R)}}\right). \text{(C4)}$$

## APPENDIX D: SCATTERING PROPERTIES AND FURTHER EXAMPLES

To calculate the CD and $g$ data shown in Figs. 2 and 4 of the main text, as well as the associated scattering matrix eigenvalues, one needs to evaluate transmission, reflection and absorption for the corresponding systems. Here we evaluate and present the reflection and transmission amplitudes as well as the total absorption $(A^{(L)}_+ + A^{(L)}_-)$, $(A^{(R)}_+ + A^{(R)}_-)$ and absorption difference (Circular Dichroism: $(A_+ - A_-)$ - side independent) when RCP/+ and LCP/- waves impinge at the left and right side of the three-layer systems of Fig. 2 of the main text, and another related system. Furthermore, we plot the eigenvalues of the scattering matrices $S^{(1)}$ and $S^{(2)}$ (Eqs. (C2) and (C4)) for the same systems. We calculate the above parameters for a range of frequencies much wider than that of Figs. 2 and 4, to illustrate that the high achievable values of the $CD$ and $g$ are not restricted to the range of Figs. 2 and 4 but are realizable in many different frequency ranges.

The first system considered is the one with $PT$-symmetric gain-loss layers (see Fig. 1 of the main text), with parameters those of Fig. 2(b), i.e. gain–loss slabs of thickness $d = 2.5\ \mu m$ and refractive index $n_{G/L} = 3 \mp 0.04i$, and chiral layer of thickness $d = 10\ nm$, chirality parameter $\kappa = \pm 5(10^{-4} + 10^{-5}i)$ and non-chiral refractive index $n_C = 1.33 + 0.01i$. In Fig. 8, left column and first panel, we show the transmission and reflection power coefficients for RCP and LCP waves incident from both sides of the system. In the second and third panels of Fig. 8-left column we calculate the total absorption and the absorption difference (CD), respectively, in the same frequency range. We observe a broad frequency range where we have multiple regions with total absorption close to zero; all these points coincide with peaks of the dissymmetry factor $g$. Regarding the absorption difference, the higher peaks here appear above the exceptional point (in the broken $PT$-phase), where the scattering matrix eigenvalues of case I (see Appendix C) diverge (see fourth panel of Fig. 8, left-column). These peaks indicate simultaneous coherent-perfect-absorption and lasing, which is one of the most exotic features of the $PT$-symmetric systems. Finally, in the last panel of Fig. 8-left column we plot the scattering matrix $S^{(2)}$ eigenvalues (case II of the appendix C) which give us the multiple positions of anisotropic transmission resonances (ATRs; points at which the eigenvalues turn from unimodular to non-unimodular and vice versa) where the absorption coefficients approach zero and maximization of $g$ takes place.

The second system considered (see second/middle column of Fig. 8) is the general non-Hermitian system of Fig. 2(c) of the main text. Here the gain layer has thickness $d = 2\ \mu m$ and refractive index $n_G = 2 - 0.05i$, the loss layer has thickness $d = 3\ \mu m$ and refractive index $n_L = 3 + 0.04i$ and the chiral layer has the same parameters as in the previous case (of the *PT*-symmetric gain-loss layers). The transmitted and reflected power for RCP and LCP waves incident from both system sides are illustrated in the middle column of Fig. 8-first panel. In the second and third panels we show the total absorption and the absorption difference (CD), respectively, in the same frequency range. Comparing with the *PT*-symmetric case, we observe a similar behaviour but a shift in frequency, due to the different gain-loss fraction. This illustrates the possibility to fully control the CD by tuning the gain-loss fraction; moreover, it indicates a flexibility and freedom in experimental realization and validation. The non *PT*-symmetry of the current system is also demonstrated by calculating the eigenvalues of scattering matrix $S^{(1)}$ (middle column, fourth panel) which are not unimodular in any frequency region.

The last system considered here has the same gain-loss configuration as in the previous case, but now the chiral layer is attached to the loss and not to the gain layer, as depicted in the top inset of the right column of Fig. 8. Examining the transmission, reflection and absorption results we observe that the position of the resonances is practically the same as in the previous case (of middle column). However, there is a quite significant difference in the CD peak amplitudes, which is connected to the different chiral-loss impedance involved here compared to the chiral-gain one of the previous case and to the reflection asymmetry of the gain-loss bilayer, leading to different field amplitudes and thus different absorption in the chiral layer region.

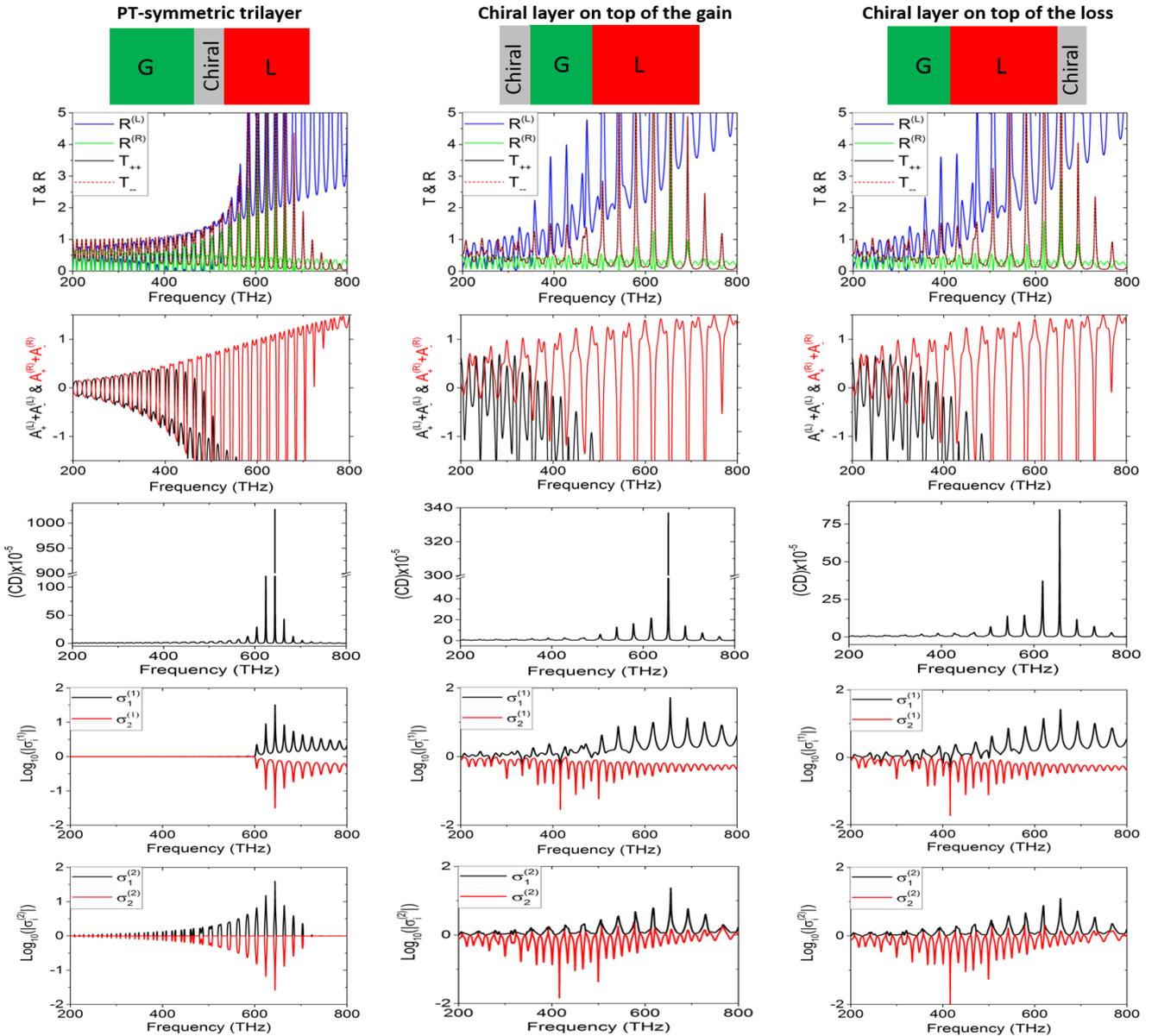

Figure 8. Transmission and reflection power coefficients ($T_{ij}=|t_{ij}|^2$, $R_{ij}=|r_{ij}|^2$, respectively, $i,j=\{+,-\}$) (first row), total absorption ($A_++A_-$) (second row), absorption difference ($A_+-A_-$) (third row), and scattering matrix eigenvalues ($\sigma$) for the scattering matrix $S^{(1)}$ (fourth row) and $S^{(2)}$ (fifth row), for RCP/+ and LCP/- circularly polarized waves incident on the systems shown in the top insets from both the left and right side (the side is indicated by the superscripts ($L$) and ($R$), respectively). Left column: system with *PT*-symmetric gain-loss layers; middle column: chiral-gain-loss general non-Hermitian system; right column: gain-loss-chiral non-Hermitian system. The material parameters and the layer thicknesses are mentioned in the text.

## APPENDIX E: STRONG CHIRAL DISSYMMETRY AT ACCIDENTAL FLUX-CONSERVING POINTS

A particularly interesting feature of *PT*-symmetric optical systems is the *accidental flux-conserving points*, corresponding to $R^{(L)} = R^{(R)}$. If $T \leq 1$, it has been shown [35] that at these points the conservation relation $T + R = 1$ of the Hermitian systems holds, although the system is non-Hermitian, resulting to zero absorption. Thus at such points, due to the vanishing of the total absorption, one can have high values of the dissymmetry factor even for very weak CD and for incidence from either side of the system. An even more intriguing feature of *PT*-systems is the possibility of $R^{(L)} = R^{(R)} = 0$, $T = 1$, i.e. an accidental flux conserving point coinciding with an anisotropic transmission resonance; such points are called points of *"double accidental degeneracy"* [35] and need special care for their engineering. At such points the dissymmetry factor can be maximized in our three-layer *PT*-systems. As an example, we calculate the dissymmetry factor for the system of Fig. 8 (left column) in a frequency region close to a point of "quasi" double accidental degeneracy. The dissymmetry factor for circularly polarized waves incident from both sides and for positive/negative chirality parameters are depicted in Figs. 9 (b) & (c). We observe strong dissymmetry factors $g^{(L)} \cong 0.24$ and $g^{(R)} \cong 0.12$ (one order of magnitude larger than the dissymmetry factors of Fig. 4) at frequency close to 119.8 THz, which is a point of "quasi" double accidental degeneracy, as shown in Fig. 9 (a) (we note again here that the presence of the chiral layer perturbs the *PT*-symmetric character of the gain-loss layers, making the three-layer system to be not fully *PT*-symmetric). Closing, we should note that the frequency region around 119.8 THz is not the only region where accidental flux conservation occurs in our system; there are multiple points/regions with this property [35].

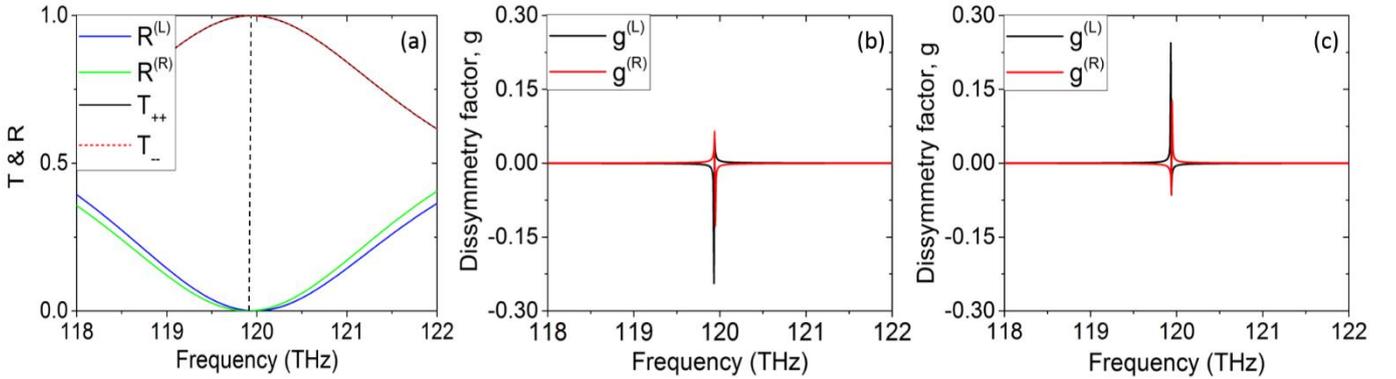

Figure 9. Panel (a): Transmission ($T$) and reflection ($R$) power coefficients for the system of Fig. 8-left-column; the dashed vertical line marks a point of "quasi" *double accidental digeneracy*. Panel (b): Dissymmetry factor ($g$) calculation for the system of panel (a) for chiral layer with $\kappa = +5(10^{-4} + 10^{-5}i)$. Panel (c): The same as in panel (b) for $\kappa = -5(10^{-4} + 10^{-5}i)$. The black lines correspond to waves incident from the left side of the system (marked by the superscript ($L$)) while the red lines from the right side (marked by the superscript ($R$)). The subscripts +/- indicate right/left circularly polarized waves.

## APPENDIX F: ABSORPTION, CIRCULAR DICHROISM AND DISYMMETRY FACTOR FOR A THIN CHIRAL LAYER - DERIVATION OF EQ. (1)

We start with Maxwell's equations and the constitutive relations for a (Pasteur) chiral medium:

$$\nabla \times \boldsymbol{E}(\boldsymbol{r},t) = -\frac{\partial \boldsymbol{B}(\boldsymbol{r},t)}{dt} \tag{F1}$$

$$\nabla \times \boldsymbol{H}(\boldsymbol{r},t) = \frac{\partial \boldsymbol{D}(\boldsymbol{r},t)}{dt} \tag{F2}$$

$$\boldsymbol{D}(\boldsymbol{r},t) = \varepsilon \boldsymbol{E}(\boldsymbol{r},t) + \frac{i\kappa}{c} \boldsymbol{H}(\boldsymbol{r},t) \tag{F3}$$

$$\boldsymbol{B}(\boldsymbol{r},t) = \mu \boldsymbol{H}(\boldsymbol{r},t) - \frac{i\kappa}{c} \boldsymbol{E}(\boldsymbol{r},t) \tag{F4}$$

where $\varepsilon = \varepsilon_r \varepsilon_0$, $\mu = \mu_r \mu_0$ are the electric permittivity and magnetic permeability, respectively, and $\kappa$ is the (Pasteur) chirality parameter. From classical electrodynamics we know that electromagnetic power density ($P$) is defined as the flux of the Poynting's vector $\boldsymbol{S}(\boldsymbol{r},t)$ which describes the flow of energy per unit time per unit area (J/sm$^2$ in SI units).

For complex time-harmonic electromagnetic fields of the form $e^{-i\omega t}$, the time averaged $P$ (over a period $T$) can be written as

$$<P>_T = [\nabla \cdot \mathbf{S}(r)] = \frac{1}{2}\text{Re}[\nabla \cdot (\mathbf{E}(r) \times \mathbf{H}^*(r))] \quad (F5)$$

Inserting Maxwell's Eqs. (F1) and (F2) into the expression $\nabla \cdot (\mathbf{E}(r,t) \times \mathbf{H}^*(r,t))$, taking into account the vector identity $\nabla \cdot (\mathbf{A} \times \mathbf{B}) = \mathbf{B} \cdot (\nabla \times \mathbf{A}) - \mathbf{A} \cdot (\nabla \times \mathbf{B})$ and assuming $\varepsilon = \varepsilon' + i\varepsilon''$, $\mu = \mu' + i\mu''$ and $\kappa = \kappa' + i\kappa''$ we can write

$$[\mathbf{H}^* \cdot (\nabla \times \mathbf{E}) - \mathbf{E} \cdot (\nabla \times \mathbf{H}^*)]$$
$$= \left[\mathbf{H}^* \cdot \left(-\frac{\partial \mathbf{B}}{\partial t}\right) - \mathbf{E} \cdot \left(\frac{\partial \mathbf{D}^*}{\partial t}\right)\right]$$
$$= [\mathbf{H}^* \cdot (i\omega \mathbf{B}) - \mathbf{E} \cdot (i\omega \mathbf{D}^*)] = [-i\omega(\mathbf{E} \cdot \mathbf{D}^* - \mathbf{B} \cdot \mathbf{H}^*)]$$
$$= -i\omega\left[\mathbf{E} \cdot \left(\varepsilon^* \mathbf{E}^* - \frac{i\kappa^*}{c}\mathbf{H}^*\right) - \left(\mu \mathbf{H} - \frac{i\kappa}{c}\mathbf{E}\right) \cdot \mathbf{H}^*\right]$$
$$= -i\omega\left[\varepsilon^*|\mathbf{E}|^2 - \mu|\mathbf{H}|^2 + \frac{i(\kappa - \kappa^*)}{c}\mathbf{E} \cdot \mathbf{H}^*\right]$$
$$= -i\omega\left[(\varepsilon' - i\varepsilon'')|\mathbf{E}|^2 - (\mu' + i\mu'')|\mathbf{H}|^2 - \frac{2\kappa''}{c}(\text{Re}(\mathbf{E} \cdot \mathbf{H}^*) + i\text{Im}(\mathbf{E} \cdot \mathbf{H}^*))\right] \quad (F6)$$

Therefore, from Eqs. (F5) and (F6), we can write

$$<P>_T = -\frac{\omega}{2}\left(\varepsilon''|\mathbf{E}|^2 + \mu''|\mathbf{H}|^2 - \frac{2\kappa''}{c}\text{Im}(\mathbf{E} \cdot \mathbf{H}^*)\right) \quad (F7)$$

Defining optical chirality [21] as

$$C = -\frac{\omega}{2c^2}\text{Im}(\mathbf{E} \cdot \mathbf{H}^*) \quad (F8)$$

Eq. (F7) becomes

$$<P>_T = -\frac{\omega}{2}(\varepsilon''|\mathbf{E}|^2 + \mu''|\mathbf{H}|^2) + 2c\kappa''C \quad (F9)$$

For loss media $\text{Re}[\nabla \mathbf{S}(r,t)] < 0$, hence the absorbed power can be computed by integrating the time-averaged power density, Eq. (F9), over the entire volume of the chiral layer. Assuming RCP/+ or LCP/- waves, we can write

$$A_\pm = \int \left[\frac{\omega}{2}(\varepsilon''|\mathbf{E}_\pm|^2 + \mu''|\mathbf{H}_\pm|^2) + 2c\kappa'' C_\pm\right] dV. \quad (F10)$$

In the case in which the chiral layer is very thin ($k_\pm d \ll 1$) or the chirality ($\kappa$) is very weak, then $|\mathbf{E}_+|^2 \approx |\mathbf{E}_-|^2 \cong |\mathbf{E}|^2$, $|\mathbf{H}_+|^2 \approx |\mathbf{H}_-|^2 \cong |\mathbf{H}|^2$ and $C_+ = -C_- \cong C$. Then, the circular dichroism, defined as: $= A_+ - A_-$, becomes

$$CD = 2c\kappa'' \int [C_+ - C_-] dV = 4c\kappa'' \int C \, dV \quad (F11)$$

and the total absorption of the system as

$$A_+ + A_- = \omega \varepsilon'' \int |\mathbf{E}|^2 \, dV + \omega\mu'' \int |\mathbf{H}|^2 \, dV + 4c\kappa'' \int C \, dV \quad (F12)$$

Combining Eqs (F11) and (F12), we can calculate the dissymmetry factor $g$ as following

$$g = \frac{2CD}{A_+ + A_-} = \frac{8c\kappa'' \int \mathbf{C} \, dV}{\omega(\varepsilon'' \int[|\mathbf{E}|^2] \, dV + \mu'' \int[|\mathbf{H}|^2] \, dV)} \quad (F13)$$

* katsantonis@iesl.forth.gr
‡ kafesaki@iesl.forth.gr